\documentclass[11pt, a4paper]{article}
\pdfoutput=1

\normalsize
\usepackage{amsmath,amssymb}
\usepackage{epsfig}
\usepackage{hyperref}
\usepackage{color}
\usepackage{cite}
\usepackage{multirow}
\usepackage{slashed}

\usepackage[font={small}]{caption}   

\definecolor{myred}{rgb}{0.7, 0, 0}
\definecolor{myblue}{rgb}{0, 0, 0.7}
\definecolor{mygreen}{rgb}{0.04, 0.7, 0.5}

\hypersetup{colorlinks,citecolor=myblue,linkcolor=myblue,urlcolor=myblue,linktocpage=true}

\newcommand{\hhref}[1]{\href{http://arxiv.org/abs/#1}{arXiv:#1}}

\newcommand{\be}{\begin{equation}}
\newcommand{\ee}{\end{equation}}
\newcommand{\bea}{\begin{eqnarray}}
\newcommand{\eea}{\end{eqnarray}}

\def\SO{\textrm{SO}}
\def\SU{\textrm{SU}}
\def\U{\textrm{U}}

\numberwithin{equation}{section}

\title{
\vspace{-2cm}
\begin{flushright}
\small{DESY 17-197}
\end{flushright}
\vspace{3cm}
\bf \LARGE
Probing light top partners with CP violation
\vspace{.2cm}}
\date{}
\author{
{\large Giuliano Panico$^{a}$, Marc Riembau$^{a,b}$, Thibaud Vantalon$^{a,b}$}\\
[10mm]
\normalsize\itshape $^a$ IFAE and BIST, Universitat Aut\`onoma de Barcelona, E-08193~Bellaterra,~Barcelona,~Spain\\
\normalsize\itshape $^b$ DESY, Notkestra{\ss}e 85, D-22607 Hamburg, Germany\\
}

\begin{document}
\maketitle
\begin{abstract}
\medskip
\noindent
We investigate CP-violating effects induced by light top partners in composite Higgs theories. We find that
sizable contributions to the dipole moments of the light SM quarks and leptons are generically generated at the
two-loop level through Barr--Zee-type diagrams. The present constraints on the electron and neutron electric dipole
moments translate into bounds on top partner masses of order $\textit{few}\;$TeV and are competitive with the
reach of LHC direct searches. Interestingly, we find that CP-violation effects are sensitive to the same operators
that control top partner single production.
Near-future improvements in the determination of the electron dipole moment
will extend the reach on top partner masses beyond the $5 - 10\;$TeV range.
\end{abstract}

\newpage

\tableofcontents


\section{Introduction}

An appealing solution to the naturalness problem is based on the idea that the Higgs boson is not an elementary state, but rather
a composite object coming from some new strongly-coupled dynamics at the TeV scale. This idea reached nowadays a quite compelling
embodiment, which is denoted as ``composite Higgs'' (CH) scenario.\footnote{See refs.~\cite{Contino:2010rs,Bellazzini:2014yua,Panico:2015jxa} for extensive reviews.}
Its main assumption is the identification of the Higgs with
a pseudo-Nambu--Goldstone boson~\cite{Kaplan:1983fs}, which, in minimal realizations, is associated to an
$\SO(5) \rightarrow \SO(4)$ symmetry-breaking pattern~\cite{Agashe:2004rs}.
An additional, fundamental ingredient is the generation of fermion masses through the partial-compositeness mechanism~\cite{Kaplan:1991dc}. The latter hypothesis is necessary to keep under control dangerously large flavor-breaking
effects and is strictly needed at least for the top quark sector.

An important consequence of partial compositeness is the presence of composite partners of the Standard Model (SM) fermions.
Among them, the partners of the top
play the most important role: besides controlling the generation
of the top mass, they also govern the leading contributions to the radiatively-induced Higgs potential~\cite{Contino:2006qr,Matsedonskyi:2012ym,Marzocca:2012zn}.
For this reason the top partners are directly connected with the amount of fine tuning and must be relatively light (around the TeV scale)
to ensure that naturalness is preserved~\cite{Panico:2012uw}.

The presence of light top partners has deep consequences for the phenomenology of CH models. First of all, being charged
under QCD, they have sizable production cross sections at hadron colliders, hence constituting one of the privileged
ways to directly test the CH paradigm at the LHC. The bounds are nowadays surpassing $1\;$TeV
(see for instance the constraints from pair production of charge-$5/3$ partners~\cite{Aaboud:2017zfn,CMS:2017wwc}), thus starting
to put some pressure on the natural parameter space of the models.

Light top partners give also rise to sizable corrections to precision observables, which can be used as powerful indirect probes
of the composite dynamics. For instance, large effects are expected in electroweak precision measurements,
such as the $S$ and $T$ parameters and the $Z$ coupling to the bottom quark. In this case the tight experimental constraints
translate into exclusions on the top partner masses around the TeV scale~\cite{Anastasiou:2009rv,Grojean:2013qca,Carena:2014ria}, which are competitive
with the ones from direct searches.

In this paper we will focus on another interesting effect due to light top partners, namely
the generation of sizable contributions to flavor physics, in particular to CP-violating observables.
These effects are due to the presence of additional complex phases in the top partners interactions.
Such phases are expected in generic composite Higgs scenarios. Complex parameters can in fact be present in the
composite sector interactions if CP-violation is allowed. Furthermore, even if the strongly-coupled dynamics is assumed
to be CP preserving, complex mixings of the elementary SM fermions
with the composite sector are still needed in many models to generate the non-trivial phase of the CKM matrix.
For instance this is the case in scenarios in which the left-handed top field is mixed with multiple composite operators.
Examples of such models are the minimal MCHM$_5$ constructions~\cite{Agashe:2004rs}.

Among the possible CP-violating effects, some of the most relevant ones are the generation of dipole moments
for the light leptons and quarks. Light top partners generically induce contributions to dipole operators
at two-loop level through Barr--Zee-type diagrams~\cite{Barr:1990vd}.\footnote{Additional contributions
can arise at the one-loop level in specific flavor set-ups, such as the ``anarchic'' scenario~\cite{Konig:2014iqa}.
They are however absent in other flavor constructions. We will discuss these aspects later on.}
Additional two-loop contributions are also generated for the gluonic Weinberg operator~\cite{Weinberg:1989dx}.
All these effects arise from the presence of CP-violating Higgs interactions involving the top and its partners.
As we will see, in a large class of models, the main contributions come from derivative Higgs interactions induced by
the non-linear Goldstone structure.\footnote{Analogous effects due to effective CP-violating Higgs interactions,
including anomalous top and bottom Yukawa couplings, have been studied in the context of the SM effective field
theory~\cite{Brod:2013cka,Kopp:2014rva}.}

The Barr--Zee effects and the Weinberg operator, in turn, give rise to sizable corrections to the
electron~\cite{Baron:2013eja,Cairncross:2017fip}, neutron~\cite{Baker:2006ts} and diamagnetic atoms~\cite{Graner:2016ses} electric dipole moments (EDM's). All these effects are tightly constrained by the present data,
moreover the experimental sensitivity is expected to increase by more than one order of magnitude in the near future~\cite{Picker:2016ygp,Doyle,Cairncross:2017fip}. As we will see, the present bounds allow to probe top
partners masses of order $\textit{few}\;$TeV and can be competitive with the direct LHC searches.
The future improvements in the EDM experiments will push the exclusions beyond the $10\;$TeV scale, arguably making
these indirect searches the most sensitive probes of top partners.

For our analysis we adopt the effective parametrizations developed in ref.~\cite{DeSimone:2012fs} and already used in the investigation
of the bounds coming from electroweak precision measurements~\cite{Grojean:2013qca}. This framework allows for a model-independent
description of the Higgs dynamics (including the whole non-linear Goldstone structure) and of the relevant composite resonances.
As we will see, top partners contributions to the dipole operators are saturated by infrared (IR) effects. The leading corrections
come from the lightest composite states and can be fully captured by the effective framework. IR saturation is instead not present
for the contributions to the Weinberg operator, therefore, we expect non-negligible ultraviolet (UV) corrections to be present.
The UV contributions, however, are expected to be independent of the IR effects and therefore should not lead to cancellations.
The light top partners contributions can thus be interpreted as a lower estimate of the full CP-violating contributions and can be safely
used to derive robust constraints.

It must be stressed that, depending on the specific flavor structure, additional contributions to flavor-violating and CP-violating
observables can be present. Typical effects can arise from partners of the light-generation SM fermions as well as
from heavy vector resonances with electroweak or QCD quantum numbers.
All these effects are generically expected in ``anarchic partial compositeness'' scenarios~\cite{anarchic} and lead to additional constraints
on the composite dynamics~\cite{anarchic_bounds,Agashe:2009di,KerenZur:2012fr,Panico:2015jxa}.
Focussing first of all on the quark sector, strong bounds on the resonance masses, of order $5 - 10\;$TeV, come from $\Delta F = 2$
observables, in particular $s \rightarrow d$ transitions that can be tested in Kaon physics. One-loop contributions to $\Delta F = 1$
and CP-violating observables, for instance the neutron EDM, are also induced by partners of the light SM quarks.
Contributions of comparable size can also be induced by the top partners due to the presence of relatively large mixing angles
with the light SM fermions. The current constraints on $\Delta F = 1$ transitions and on the neutron EDM translate into bounds
on the resonance masses of order $\textit{few}\;$TeV.
If the ``anarchic'' construction is naively extended to the lepton sector, more dangerous flavor effects arise~\cite{KerenZur:2012fr}.
In this case large one-loop contributions to the electron EDM and to $\mu \rightarrow e \gamma$ transitions are generated,
which can be compatible with the present experimental bounds only if the scale of new physics is of order $50-100\;$TeV.
In this scenario the two-loop contributions from top partners are clearly subdominant. Due to the extremely strong bounds, however,
we find the naive ``anarchic partial compositeness'' scenario too fine-tuned to be considered as a fully satisfactory set-up.

Models featuring flavor symmetries can significantly help in reducing the experimental constraints. Several scenarios based
on $\U(3)$~\cite{Redi:2011zi} or $\U(2)$~\cite{Barbieri:2012uh} symmetries in the quark sector have been proposed.
In these cases leading contributions to flavor-violating and CP-violating observables are reduced and a compositeness scale
around $\textit{few}\;$TeV is still allowed. The flavor symmetry structure can also be extended to the lepton sector~\cite{Redi:2013pga},
thus keeping under control the one-loop contributions to the electron EDM and $\mu \rightarrow e \gamma$ transitions.
In these scenarios the two-loop CP-violating effects we consider in this paper can still be present and can give significant
bounds on the mass of the top partners. Notice that additional phenomenological handles
are typically present in these models due to the sizable amount of compositeness of the light generation fermions~\cite{Delaunay:2010dw}.

Another appealing flavor scenario, which has been recently proposed in the literature, is based on a departure from the classical
partial compositeness paradigm for the light SM fermions~\cite{Vecchi:2012fv,Panico:2016ull}.
In these models only the top quark (or at most the third generation fermions) are assumed to be partially composite objects
at the TeV scale, while the Yukawa couplings of the light SM fermions are generated by a dynamical mechanism at much
higher energy scales. This construction leads to an effective minimal flavor violation structure and efficiently reduces all flavor-violating
and CP-violating effects, most noticeably in the lepton sector~\cite{Panico:2016ull}.
The bounds on the masses of the composite states are lowered
to the $\textit{few}\;$TeV range, thus allowing for natural models with a small amount of fine-tuning.
In these scenarios CP-violating effects from top partners are expected to play
a major role and can lead to the strongest bounds on the compositeness scale.

The paper is organized as follows. In sec.~\ref{sec:simpl_model} we analyze the generation of
CP-violating dipole moments induced by light top partners in a simplified set-up with only one composite fermion
multiplet. We show that dipole operators are mainly due to running effects coming from effective contact Higgs interactions,
and we derive full analytical expressions for the CP-violating effects. Afterwards we discuss the bounds
on the top partner masses coming from electron, neutron and mercury EDM measurements
and we compare them with the
exclusions from direct searches at the LHC and future colliders. In sec.~\ref{sec_non_minimal_models} we extend
the analysis to non-minimal scenarios, investigating the effects due to the presence of additional light top partner
multiplets. Finally we conclude in sec.~\ref{sec:conclusions}.

\section{CP violation from top partners}\label{sec:simpl_model}

To discuss the general features of CP violation in composite models, and in particular the generation of electron and neutron
EDM's, in this section we focus on a simplified model containing only one multiplet of top partners.
As we will see, this set-up retains all the main features of more complex models, but allows us to
obtain a simpler qualitative and quantitative understanding of CP-violating effects.
Non-minimal scenarios with multiple top partners will be discussed in sec.~\ref{sec_non_minimal_models}.

For definiteness, we restrict our attention to the class of minimal composite Higgs models
based on the global symmetry breaking pattern $\SO(5) \rightarrow \SO(4)$~\cite{Agashe:2004rs}.\footnote{In order
to accommodate the correct fermion hypercharges an additional $\U(1)_X$ global Abelian subgroup is needed
(see for instance ref.~\cite{Panico:2015jxa}).}
This pattern gives rise to only one Goldstone Higgs doublet and preserves an
$\SO(3)_c$ custodial symmetry, which helps in keeping under control corrections to the electroweak precision parameters.
Motivated by fine-tuning considerations (see refs.~\cite{Panico:2012uw,Matsedonskyi:2015dns}), we assume
that the $\SU(2)_L$ doublet $q_L = (t_L, b_L)$ is linearly mixed with composite operators in the ${\bf 14}$
representation of $\SO(5)$. The right-handed top component is instead identified with a fully composite
chiral singlet coming from the strongly-coupled dynamics. 
This scenario is usually dubbed $\bf 14+1$ model~\cite{DeSimone:2012fs,Panico:2012uw}.

The possible quantum numbers of the top partners are determined by the unbroken $\SO(4)$ symmetry. From the
decomposition ${\bf 14} = {\bf 9} \oplus {\bf 4} \oplus {\bf 1}$, one infers that the partners can fill the nineplet, fourplet
or singlet representations of $\SO(4)$. As we will see, the main CP-violating effects typically arise form the lightest
top partner multiplet. Restricting the analysis to a limited set of partners is thus usually a good approximation.
For simplicity in this section we will consider a scenario in which the lightest partners transform
in the fourplet representation.

The most general leading-order effective action for the SM quarks and a light composite fourplet $\psi_4$
can be written in the CCWZ framework~\cite{ccwz} (see ref.~\cite{Panico:2015jxa} for an in-depth review of the
formalism) as
\begin{eqnarray}
{\cal L} &=& i \overline q_L \slashed D q_L + i \overline t_R \slashed D t_R
+ i\overline \psi_4 (\slashed D - i \slashed e) \psi_4
- \left(m_4 \overline \psi_{4L} \psi_{4R} + {\rm h.c.}\right)\nonumber\\
&& +\, \left(-i\, c_t \overline \psi_{4R}^i \gamma^\mu d_\mu^i t_{R}
+ \frac{y_{Lt}}{2} f (U^t \overline q_L^{\bf 14} U)_{55} t_R + y_{L4} f (U^t \overline q_L^{\bf 14} U)_{i5} \psi_{4R}^i  + {\rm h.c.}\right)\,.
\label{eq:Lcomp_14+1}
\end{eqnarray}
In the above formula $q_L^{\bf 14}$ denotes the embedding of the $q_L$ doublet into the representation $\bf 14$,
explicitly given by
\begin{equation}
q^{\bf 14}_L = \frac{1}{\sqrt{2}}\left(
\begin{array}{ccccc}
0 & 0 & 0 & 0 & -i b_L \\
0 & 0 & 0 & 0 & -b_L \\
0 & 0 & 0 & 0 & -i t_L \\
0 & 0 & 0 & 0 & t_L \\
- i b_L & -b_L & -i t_L & t_L & 0
\end{array}
\right)\,.
\end{equation}
The Goldstone Higgs components $\Pi_i$, in the real fourplet notation, are encoded in the matrix
\begin{equation}
U = \exp\left[i \frac{\sqrt{2}}{f} \Pi_i \widehat T^i\right]\,,
\end{equation}
where $f$ is the Goldstone decay constant and $\widehat T^i$ ($i = 1,\ldots,4$) are the generators of the
$\SO(5)/\SO(4)$ coset.
In the first line of eq.~(\ref{eq:Lcomp_14+1}), $D_\mu$ denotes the usual covariant derivative containing the SM
gauge fields. The $d_\mu$ and $e_\mu$ symbols denote the CCWZ operators, defined as
\begin{equation}
U^t [A_\mu + i \partial_\mu] U = e_\mu^a T^a + d_\mu^i \widehat T^i\,,
\end{equation}
with $T^a$ ($a = 1,\ldots,6$) the $\SO(4)$ generators and $A_\mu$ the SM gauge fields rewritten in an $\SO(5)$ notation.

We can now easily identify possible sources of CP violation. The effective Lagrangian in eq.~(\ref{eq:Lcomp_14+1})
contains four free parameters, namely $m_4$, $y_{Lt}$, $y_{L4}$ and $c_t$. In general all of them are complex.
By using chiral rotations, however, three parameters can be made real, so that only one physical complex phase
is present in the model.
It can be easily seen that $m_4$ can be always made real by a phase redefinition of $\psi_{4L}$.
This redefinition does not affect the other parameters. The complex phases of the remaining three parameters are instead
connected. The elementary-composite mixing parameters $y_{Lt}$ and $y_{L4}$ can be made real through
phase rotations of $t_R$ and $\psi_{4R}$, shifting all the complex phases into $c_t$.
CP-violating effects are thus controlled by the complex phase of the combination $c_t y_{Lt}^* y_{L4}$.

Complex values of the elementary-composite mixing parameters can in general be present even if CP invariance is imposed
in the composite sector (so that $m_4$ and $c_t$ are real). This is the case, for instance if the $q_L$ doublet is coupled
with two composite operators in the UV, eg. with an operator ${\cal O}_L$ corresponding to the fourplet partners and with another
${\cal O}_R$ corresponding to the composite $t_R$. It is however also possible that a single dominant mixing with ${\cal O}_L$
is present. In this case one expects $y_{Lt}$ and $y_{L4}$ to have the same complex phase, thus avoiding CP-violation from top partners
if the composite sector preserves CP.

It is also interesting to notice that, in the set-up we are considering, CP-violation is unavoidably linked to the presence
of $d_\mu$-interaction operators. If the term $-i\, c_t \overline \psi_{4R}^i \gamma^\mu d_\mu^i t_{R}$ is not present
in the effective Lagrangian, CP is preserved. We will see in section~\ref{sec:2-site}, that a similar result is also valid in more
generic models with additional top partners and multiple physical complex phases.

\subsection{Electron EDM}\label{sec:e-EDM}

\begin{figure}
\centering
\includegraphics[width=.35\textwidth]{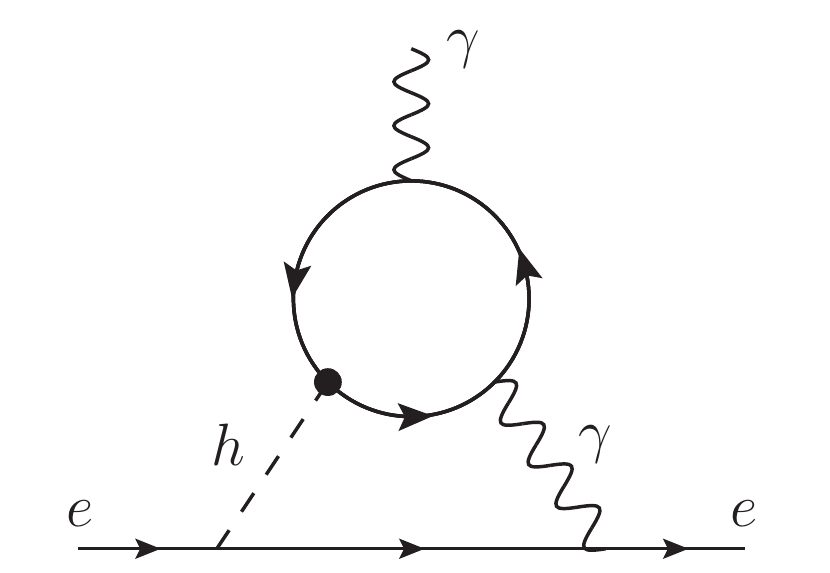}
\caption{Barr--Zee type diagram giving rise to the contribution to the electron EDM.}\label{fig:Barr-Zee}
\end{figure}

The presence of CP-violating interactions of the top and its partners can give rise to sizable contributions to
EDM's. In particular an EDM for the electron,
\begin{equation}
{\cal L}_{eff} = - d_e \frac{i}{2} \overline e \sigma^{\mu\nu} \gamma^5 e F_{\mu\nu}\,,
\end{equation}
arises at two-loop level through Barr--Zee diagrams involving CP-violating Higgs interactions~\cite{Barr:1990vd}
(see fig.~\ref{fig:Barr-Zee}). In this subsection we will investigate in detail how this effect arises and derive explicit expressions
to compute it.

To discuss the CP-violating effects it is convenient to choose a field basis in which the physical complex phase is
put into $c_t$, while the remaining parameters are real. In this basis, CP-violating Higgs couplings to
the top quark and its partners arise only from the $-i\, c_t \overline \psi_{4R}^i \gamma^\mu d_\mu^i t_{R}$ operator.
At leading order in the $v/f$ expansion, where $v \simeq 246\;$GeV denotes the Higgs vacuum expectation value, we obtain
\begin{equation}\label{eq:d_mu-couplings}
-i\, c_t \overline \psi_{4R}^i \gamma^\mu d_\mu^i t_{R} + {\rm h.c.} \supset i \frac{c_t}{f} \partial_\mu h
\left(\overline {\widehat X}_{2/3R} \gamma^\mu t_R - \overline {\widehat T}_R \gamma^\mu t_R\right) + {\rm h.c.}\,,
\end{equation}
where we used the decomposition of the $\psi_4$ fourplet into components with definite quantum numbers
under the SM group
\begin{equation}
\psi_4 = \frac{1}{\sqrt{2}}\left(
\begin{array}{c}
-i B + i X_{5/3}\\
- B - X_{5/3}\\
-i \widehat T - i \widehat X_{2/3}\\
\widehat T - \widehat X_{2/3}
\end{array}
\right)\,.
\end{equation}
The components of $\psi_4$ correspond to two $\SU(2)_L$ doublets, namely $(\widehat T, B)$
and $(X_{5/3}, \widehat X_{2/3})$, with hypercharges $1/6$ and $7/6$ respectively.

The main contributions to the electron EDM arise from Barr--Zee diagrams involving a virtual photon. Additional
corrections come from diagrams involving a virtual $Z$ boson. These contributions, however,
are proportional to the vector coupling of the $Z$ to the charged leptons, which is accidentally small in the
SM~\cite{Barr:1990vd,Brod:2013cka}. They are thus strongly suppressed and can be safely neglected.

Since the photon couplings are flavor-blind and diagonal, the most convenient way to evaluate the Barr--Zee diagrams
is to perform the computation in the mass eigenstate basis. In this way each fermionic state
gives an independent contribution to the electron EDM.  From the explicit form of the couplings in
eq.~(\ref{eq:d_mu-couplings}) it can be seen that only the charge-$2/3$ fields have CP-violating interactions
involving the Higgs, thus these states are the only ones relevant for our computation.

The spectrum of the charge-$2/3$ states is quite simple. One combination of the $\widetilde T$ and $\widetilde X_{2/3}$
fields (which we denote by $X_{2/3}$) does not mix with the elementary fields and has a mass $m_{X_{2/3}} = |m_4|$.
The orthogonal combination
\begin{equation}
T = \frac{1}{\sqrt{2 + \cos(2v/f) + \cos(4v/f)}} \left[\left(\cos(v/f) + \cos(2v/f)\right) \widehat T
+ \left(\cos(v/f) - \cos(2v/f)\right) \widehat X_{2/3}\right]\,,
\end{equation}
is mixed with the elementary top field and its mass acquires a shift controlled by the $y_{L4}$ parameter, plus
an additional subleading correction due to electroweak symmetry breaking,
\begin{equation}
m_T \simeq \sqrt{m_4^2 + y_{L4}^2 f^2}\left[1 - \frac{5}{4}\frac{y_{L4}^2 f^2}{m_4^2}\frac{v^2}{f^2} + \cdots\right]\,.
\end{equation}
The top mass is mostly determined by the $y_{Lt}$ parameter and, at leading order in the $v/f$ expansion, reads
\begin{equation}
m_{top}^2 \simeq \frac{1}{2} \frac{m_4^2}{m_4^2 + y_{L4}^2 f^2} y_{Lt}^2 v^2\,.
\end{equation}
The full spectrum of the model also includes the $X_{5/3}$ field with electric charge $5/3$ and mass
$m_{X_{5/3}} = |m_4|$ and the $B$ field with electric charge $-1/3$ and mass $m_B = \sqrt{m_4^2 + y_{L4}^2 f^2}$.
Notice that the $X_{5/3}$ and $X_{2/3}$ states are always the lightest top partners in the present set-up.

In order to compute the electron EDM, we need to determine the flavor-diagonal CP-violating couplings of the Higgs
to the fermion mass eigenstates, in particular the top, the $T$ and the $X_{2/3}$. It turns out that the $X_{2/3}$ field does not
have such coupling, as a consequence of the fact that it has no mass mixing with the elementary states.
The relevant couplings are thus given by
\begin{equation}\label{eq:CP_viol_vert}
\frac{1}{f} \partial_\mu h \left[c_{top} \overline t_R \gamma^\mu t_R + c_T \overline T_R \gamma^\mu T_R\right]\,,
\end{equation}
where, at leading order in $v/f$,
\begin{equation}\label{eq:cT_value}
c_{T} = - c_{top} = {\rm Im}\, c_t\, \sin 2 \varphi_R = \sqrt{2} v \frac{y_{L4} y_{Lt} f}{m_4^2 + y_{L4}^2 f^2}  {\rm Im}\,c_t = 2\, {\rm Im}\, c_t\, \frac{y_{L4} f}{\sqrt{m_4^2 + y_{L4}^2 f^2}} \frac{m_{top}}{m_4}\,.
\end{equation}
In the above expression $\varphi_R$ denotes the rotation angle that diagonalizes the mass matrix of the $t_R$ and
$T_R$ fields. Notice that the operators in eq.~(\ref{eq:CP_viol_vert}) are necessarily CP-odd and their coefficients
are real.

The result in eq.~(\ref{eq:cT_value}) shows that the CP-violating couplings for the top quark and the $T$ field have
opposite coefficients. This relation is exact at all orders and is a consequence of the fact
that the interactions coming from the $d_\mu$-operator in the Lagrangian~(\ref{eq:Lcomp_14+1}) are strictly off-diagonal.
The trace of the coupling matrix must therefore
vanish, so that the sum of the coefficients of the diagonal interactions in the mass eigenstate basis is aways zero.
This result can be easily generalized to scenarios with multiple top partners and with $d_\mu$ interactions that
involve both fermion chiralities. In this case the sum of the coefficients
of the CP-violating Higgs interactions over all fermions vanishes independently for each coupling chirality, namely
$\sum_i c_{i\textsc{l}} = \sum_i c_{i\textsc{r}} = 0$.

\begin{figure}
\centering
\raisebox{.5em}{\includegraphics[width=.3\textwidth]{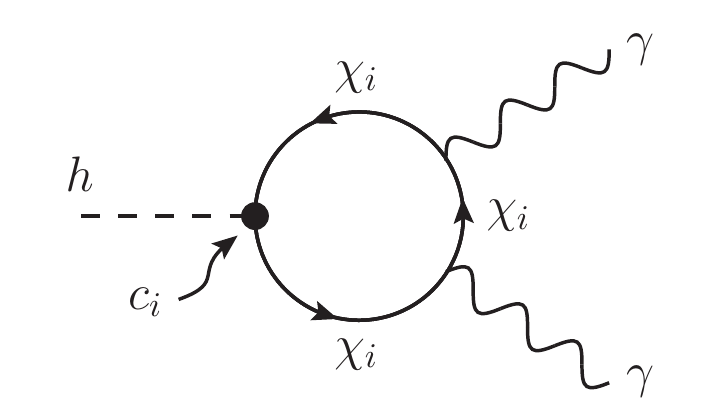}}
\hspace{3.5em}
\includegraphics[width=.3\textwidth]{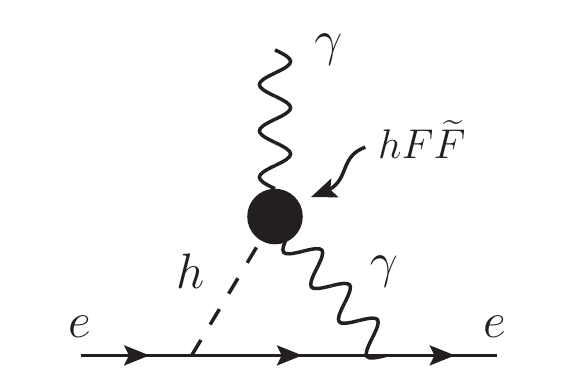}
\caption{Contribution to the electron EDM from running.}\label{fig:Running}
\end{figure}

\subsubsection{Electron EDM as a running effect}

Instead of presenting straight away the full result of the computation of the Barr--Zee diagrams, we find more instructive
to follow a simplified approach that allows us to highlight a deeper physical origin of the EDM's.
The full result will be presented in sec.~\ref{sec:full_res}.

As a first step we focus on a single fermion mass eigenstate with CP-violating interactions analogous to the ones in
eq.~(\ref{eq:CP_viol_vert}).
It is straightforward to see that such couplings give rise at one loop to CP-violating effective interactions among
the Higgs and two photons, originating from diagrams analogous to the one shown in the left panel of fig.~\ref{fig:Running}.
Parametrizing the CP-violating Higgs interactions as
\begin{equation}
{\cal L} \supset \frac{c_{i\textsc{l,r}}}{f} \, \partial_\mu h\, \overline \chi_i \gamma^\mu P_{L,R} \chi_i\,,
\end{equation}
where $P_{L,R} = (1 \mp \gamma^5)/2$ are the left and right chirality projectors,
we find that the one-loop matrix element is given by
\begin{equation}\label{eq:HGG_matrix_element}
{\cal M} = \pm i \frac{N_c}{2 \pi^2 s} e^2 Q_{f_i}^2\,  \varepsilon_{\mu\nu\rho\sigma}\, \varepsilon^\nu(\lambda_1, k_1)\, \varepsilon^\mu(\lambda_2, k_2)\,
k_1^\rho k_2^\sigma\, \frac{c_{i\textsc{l,r}}}{f} m_i^2 F(4 m_i^2/s)\,.
\end{equation}
where the $F$ function is defined as
\begin{equation}
F(\tau) = \left\{
\begin{array}{l@{\hspace{2.em}}c}
\displaystyle\frac{1}{2} \left[\log \frac{1 + \sqrt{1-\tau}}{1 - \sqrt{1 - \tau}} - i \pi\right]^2 & {\rm for}\ \tau < 1\\
\rule{0pt}{1.5em}- 2 \arcsin^2 (1/\sqrt{\tau}) & {\rm for}\ \tau \geq 1
\end{array}
\right.\,.
\end{equation}
In eq.~(\ref{eq:HGG_matrix_element}), $Q_{f_i}$ denotes the fermion electric charge (in the present set-up $Q_{f_i} =2/3$),
$k_{1,2}$ and $\varepsilon^{\mu,\nu}(\lambda_{1,2}, k_{1,2})$ are the momenta and the polarization vectors of the photons,
while $s = (k_1 + k_2)^2$ coincides with $m_h^2$ for an on-shell Higgs.

The above result can be matched onto a series of CP-violating effective operators analogous to
$(\square^n H^2) F_{\mu\nu} \widetilde F^{\mu\nu}$,
where $F$ is the photon field strength and $\widetilde F_{\mu\nu} = 1/2 \varepsilon_{\mu\nu\rho\sigma} F^{\rho\sigma}$ is the dual
field-strength tensor.
For this purpose it is convenient to expand $|{\cal M}|^2$ as a series in $s/m_i^2$. In particular,
for $4 m_i^2 > s$ we find that the first terms in the expansion are
\begin{equation}
F(4 m_i^2/s) \simeq - \frac{s}{2 m_i^2} - \frac{s^2}{24 m_i^4} + \cdots\,.
\end{equation}
The leading term matches onto the effective operator
\begin{equation}\label{eq:HGG_matched}
\mp \frac{e^2 N_c Q_{f_i}^2}{16 v \pi^2} \frac{c_{i\textsc{l,r}}}{f} H^2  F_{\mu\nu} \widetilde F^{\mu\nu}\,,
\end{equation}
while the second term in the series corresponds to an effective operator involving two additional derivatives.

At the one-loop level, the $H^2 F_{\mu\nu} \widetilde F^{\mu\nu}$ effective operator gives rise to a logarithmically divergent
diagram (see right panel of fig.~\ref{fig:Running}) that induces a running for the electron EDM operator.
The divergence, and thus the running, is eventually regulated by the Higgs mass $m_h$.
The effective operator in eq.~(\ref{eq:HGG_matched}) leads to the contribution
\begin{equation}\label{eq:EDM-running}
\frac{d_e}{e} = \mp\frac{N_c}{64\pi^4} e^2 Q_{f_i}^2 \frac{y_e}{\sqrt{2}} \frac{c_{i\textsc{l,r}}}{f} \log \frac{m^2_i}{m^2_h}\,,
\end{equation}
where $y_e$ denotes the electron Yukawa coupling.

To find the full contribution to $d_e$ in our simplified $\bf 14 + 1$ model, we need to sum over the contributions of the $T$ resonance
and of the top. In this way we find the leading logarithmically-enhanced contribution to the electron EDM
\begin{equation}\label{eq:eEDM_app}
\frac{d_e}{e} = -\frac{e^2}{48\pi^4} \frac{y_e}{\sqrt{2}} \frac{c_T}{f} \log \frac{m^2_T}{m^2_{top}}\,.
\end{equation}
We will see in sec.~\ref{sec:full_res} that this is the dominant contribution to the electron EDM,
and additional threshold effects are subleading.

A few comments are in order. Although the result in eq.~(\ref{eq:eEDM_app}) is logarithmically enhanced for large $m_T$,
its overall coefficient $c_T$ is inversely proportional to the top partner mass (see eq.~(\ref{eq:cT_value})). The overall effect
is thus dominated by the contributions coming from the lightest top partners and is largely insensitive to the UV
details of the theory.

\begin{figure}
\centering
\includegraphics[width=.4\textwidth]{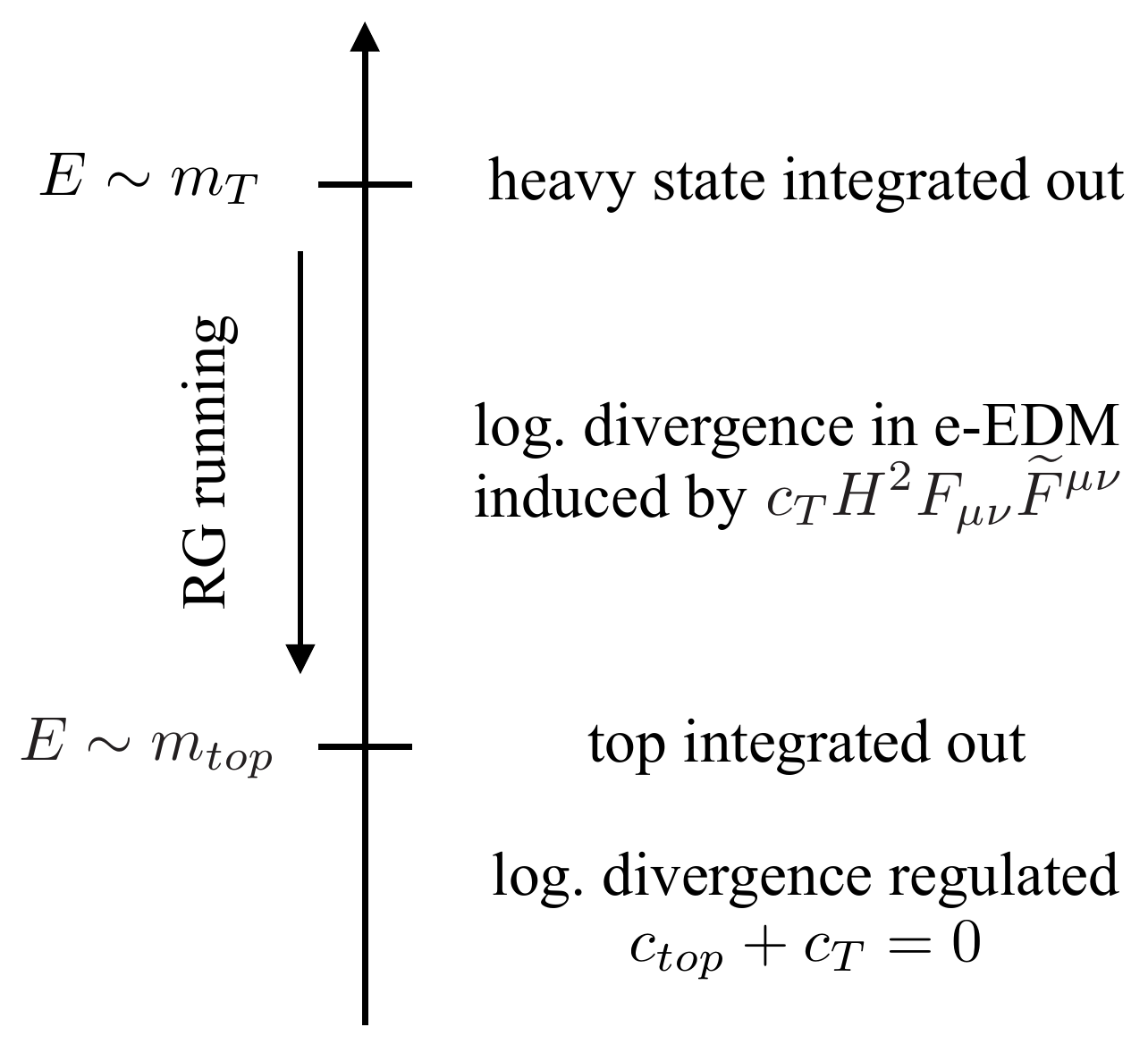}
\caption{Schematic cartoon explaining the generation of an electron EDM as a two-loop running effect due to the top partners.}\label{fig:Running_cartoon}
\end{figure}

It is also interesting to notice that the argument of the logarithm is given by the ratio of the $T$ resonance mass and the top mass,
whereas the Higgs mass that appeared in eq.~(\ref{eq:EDM-running}) is not present in the final result.
This can be understood by comparing the contributions of the $T$ and top loops to the electron EDM running.
As schematically shown in fig.~\ref{fig:Running_cartoon}, at the $m_T$ scale a contribution to the $H^2 F_{\mu\nu} \widetilde F^{\mu\nu}$ effective operator is generated, giving rise to a running for the electron EDM.
A second contribution, exactly opposite to the first one, is then generated at the top mass scale, stopping the running.
The exact compensation of the $T$ and top contributions is a consequence of the relation $c_T = - c_{top}$.

This feature is not a peculiarity of our simple set-up, but is quite generic. Since the sum of all the CP violating
coefficients $c_{i\textsc{l},\textsc{r}}$ vanishes, the total contributions to the effective operator $H^2 F_{\mu\nu} \widetilde F^{\mu\nu}$
sum up to zero and the running effects in the electron EDM are always regulated at the top mass scale.
This result has an interesting consequence for Higgs physics, since it forbids sizable CP-violating contributions
to the Higgs decay into a photon pair. Effects of this type can only come from higher-dimension operators
like $(\square^n H^2) F_{\mu\nu} \widetilde F^{\mu\nu}$, and are necessarily suppressed by additional
factors $(m^2_h/m^2_i)^{n}$. The contributions from heavy top partners are thus typically negligible, while relevant
corrections can only come from the top quark.

\subsubsection{The full result}\label{sec:full_res}

We can now present the full computation of the top partners contribution to the electron EDM.
For this purpose it is convenient to rewrite the CP-violating Higgs interactions in an equivalent form.
Integrating by parts and using the equations of motion for the fermions (or equivalently by a suitable field redefinition),
we can rewrite the interactions arising from the $d_\mu$ operators as CP-odd Yukawa couplings
\begin{equation}\label{eq:d_mu_rewriting}
\frac{c_{i\textsc{l,r}}}{f} \partial_\mu h \overline \chi_{i} \gamma^\mu P_{L,R} \chi_{i} \rightarrow \pm i \frac{c_{i\textsc{l,r}}}{f} m_i\, h\, \overline \chi_i \gamma^5 \chi_i\,.
\end{equation}
The full two-loop Barr--Zee diagram involving CP-odd top Yukawa's has been computed in refs.~\cite{Barr:1990vd,Stockinger:2006zn,Brod:2013cka}.  Using these results we find that the full two-loop contribution
to the electron EDM for a generic set of fermionic resonances is given by
\begin{equation}\label{eq:e-EDM}
\frac{d_e}{e}= 4 \frac{N_c}{f} \frac{\alpha}{(4\pi)^3} \frac{y_e}{\sqrt{2}} \sum_i Q_{f_i}^2 (c_{i\textsc{r}} - c_{i\textsc{l}}) f_1(x_i)\,,
\end{equation}
where $x_i=m_i^2/m_h^2$ and the $f_1$ function is given by
\begin{equation}
f_1(x)=\frac{2x}{\sqrt{1-4x}}\left[{\rm Li}_2\left(1-\frac{1-\sqrt{1-4x}}{2x}\right) - {\rm Li}_2\left(1-\frac{1+\sqrt{1-4x}}{2x}\right)\right]\,,
\end{equation}
with ${\rm Li}_2$ denoting the usual dilogarithm
${\rm Li}_2(x) = -\int_0^x\,du \frac{1}{u}\log (1-u)$.

To make contact with the result obtained in the previous section, we can expand the $f_1(x)$ function for large $x$
(i.e.~large fermion masses $m_i \gg m_h$), obtaining
\begin{equation}
\sum_i (c_{i\textsc{r}} - c_{i\textsc{l}}) f_1(x_i)=\,\sum_i (c_{i\textsc{r}} - c_{i\textsc{l}}) \left[\log x_i +\frac{1}{x_i}\left(\frac{5}{18}+\frac{1}{6}\log x_i\right) + \cdots\right]\,,
\end{equation}
where we used $\sum_i c_{i\textsc{r}} = \sum_i c_{i\textsc{l}} = 0$.
We can see that the leading logarithmic term exactly matches the result in eq.~(\ref{eq:EDM-running}). As expected, the subleading
terms are suppressed by powers of $m_h^2/m_i^2$ and would match the contributions from
higher-derivatives effective operators. It is interesting to notice that the subleading terms are also further suppressed by accidentally
small numerical coefficients, and are almost negligible already for the top contributions.

\subsection{CP-violating effects for the light quarks}

\begin{figure}
\centering
\includegraphics[width=.325\textwidth]{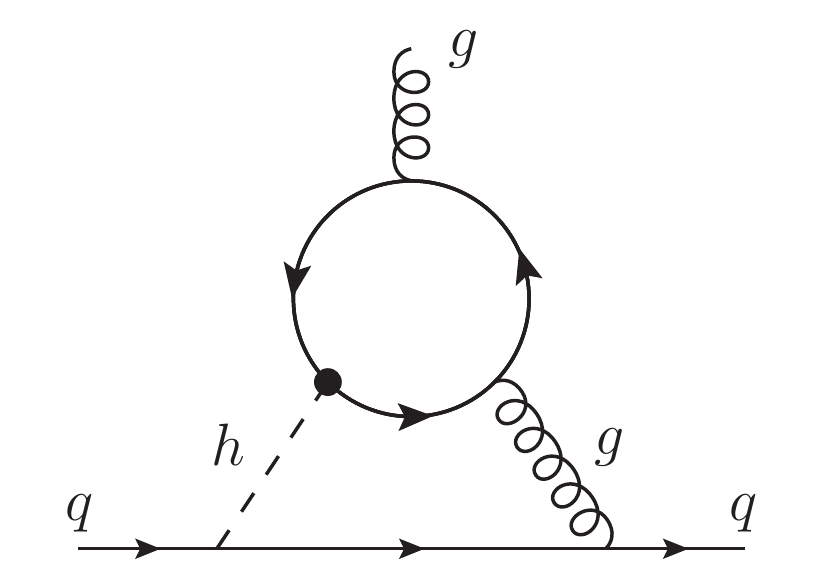}
\hspace{3.5em}
\raisebox{.5em}{\includegraphics[width=.235\textwidth]{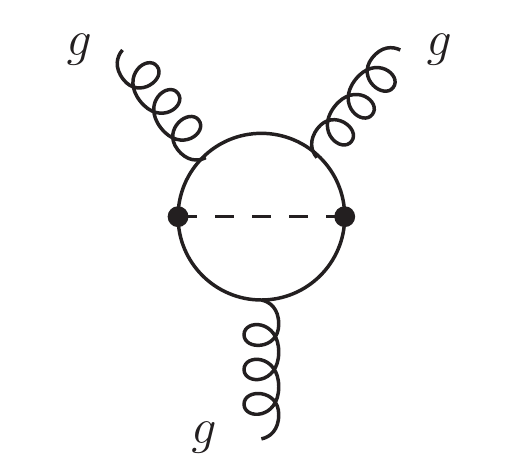}}
\caption{Two-loop diagrams giving rise to a cromoelectic dipole moment for the light quarks (left) and
to the Weinberg operator (right).}\label{fig:Weinberg_op}
\end{figure}

The anomalous top and top partner couplings with the Higgs give also rise to additional CP-violating effects. The main ones are
electric and chromoelectric dipole moments for the light quarks and a contribution to the gluonic Weinberg operator~\cite{Weinberg:1989dx}.
The light quark EDM's arise through two-loop diagrams similar to the one giving rise to the electron EDM (see fig.~\ref{fig:Barr-Zee}),
but with the electron line replaced by a quark line. The chromoelectric dipole moments (CEDM's) arise instead from
Barr--Zee-type diagrams involving gluons, as shown in the left panel of fig.~\ref{fig:Weinberg_op}. Finally the
Weinberg operator is generated by two-loop diagrams of the type shown in the right panel of fig.~\ref{fig:Weinberg_op}.
Notice that the Weinberg operator arises from diagrams that involve only the couplings of the Higgs to the top and top partners,
hence it is independent of the light quark Yukawa's.

The dipole moments of the light quarks and the Weinberg operator can be parametrized through the following effective
Lagrangian
\begin{equation}
{\cal L}_{eff} = - d_q \frac{i}{2} \overline q \sigma^{\mu\nu} \gamma^5 q F_{\mu\nu}
- \widetilde d_q \frac{i g_s}{2} \overline q \sigma^{\mu\nu} T^a \gamma^5 q G^a_{\mu\nu}
- w \frac{1}{3} f^{abc} G^a_{\mu\sigma} G^{b,\sigma}_\nu \widetilde G^{c,\mu\nu}\,,
\end{equation}
where $q = u,d$ denote the first generation quarks, $\widetilde G^{a,\mu\nu} = \frac{1}{2} \varepsilon^{\mu\nu\rho\sigma} G^a_{\rho\sigma}$ is the dual QCD field-strength tensor and $T^a$ are the color generators, normalized as ${\rm Tr}[T^a, T^b] = \delta_{ab}/2$.

The quark EDM's and CEDM's can be straightforwardly computed as we did in the previous subsection for the electron EDM.
The full results are given by
\begin{eqnarray}
d_q &=& - 4 Q_q \frac{N_c}{f} e \frac{\alpha}{(4\pi)^3} \frac{y_q}{\sqrt{2}} \sum_i Q_{f_i}^2 (c_{i\textsc{r}} - c_{i\textsc{l}}) f_1(x_i) \,,\label{eq:dq}\\
\widetilde d_q &=& - \frac{2}{f} \frac{\alpha_s}{(4\pi)^3} \frac{y_q}{\sqrt{2}} \sum_i (c_{i\textsc{r}} - c_{i\textsc{l}}) f_1(x_i) \,,\label{eq:dqtilde}
\end{eqnarray}
where $y_q$ denote the light quark Yukawa couplings.

Let us now consider the Weinberg operator. The structure of the two-loop diagram contributing to this operator
makes it sensitive to a larger set of CP-violating sources. Differently from the Barr--Zee-type contributions, the diagrams
giving rise to the Weinberg operator involve a fermion loop with two insertions of Higgs couplings. As a consequence
they receive contributions not only from the diagonal Higgs interactions, but also from the off-diagonal couplings
involving two different fermion mass eigenstates~\cite{Dicus:1989va}. 

Three sets of diagrams give rise to contributions to the Weinberg operator.
The first set includes diagrams involving a CP-even Yukawa coupling and a CP-odd derivative Higgs interaction
coming from the $d_\mu$ operator. As we already mentioned, these contributions can also come from fermion loops
involving two different fermionic mass eigenstates. In fact, in generic composite Higgs theories, including
the simplified set-up considered in this section, the Higgs couplings to the top and top partners also
have off-diagonal terms. This is true both for the Yukawa couplings and for the interactions coming from the
$d_\mu$ operator.

The second class of contributions comes from diagrams involving two Yukawa couplings. In a large class of models
the diagonal Yukawa couplings are always CP-even, in such case the contributions to the Weinberg operator can only come
from diagrams involving two off-diagonal Higgs interactions.

Diagrams in the third class involve two $d_\mu$ derivative Higgs interactions.
Since diagonal couplings of this type are necessarily CP-odd, the only contributions of this kind to the Weinberg operator come from
the off-diagonal Higgs interactions. Such interactions can have both a CP-even and a CP-odd component.

Notice that, in the model we are considering in this section, only the first class of contributions is present,
while diagrams involving two Yukawa couplings or two $d_\mu$ interactions do not give rise to CP-violating effects.
The absence of contributions induced only by the Yukawa couplings is a consequence of the fact that, through a field redefinition,
all complex phases can be removed from the mass parameters and from the mixings between the composite
resonances and the elementary states. In this basis the only CP-violating vertices come from the $d_\mu$ interactions.
Diagrams involving only $d_\mu$ couplings are instead absent since in our simplified set-up with only one light multiplet
all these interactions have the same complex phase, which cancels out in the final result. We will discuss this in detail in the following.

The contribution to the Weinberg operator coming from a set of fermions with Yukawa couplings of the form
\begin{equation}
{\cal L} = - \frac{1}{\sqrt{2}} \sum_{i,j} \overline \psi_i \left[y_{ij} + i \widetilde y_{ij} \gamma^5\right] \psi_j h\,,
\end{equation}
is given by~\cite{Dicus:1989va}
\begin{equation}\label{eq:w_diagram}
w = \frac{g_s^3}{4 (4\pi)^4} \sum_{i,j} \frac{{\rm Re} [y_{ij} \widetilde y_{ij}^*]}{m_i m_j}
f_3(x_i, x_j)\,,
\end{equation}
where the function $f_3$ is defined as
\begin{equation}
f_3(x_i, x_j) = 2 x_i x_j \int_0^1 dv \int_0^1 du \frac{u^3 v^3 (1-v)}{[x_i uv(1-v) + x_j v(1-u)
+ (1-v)(1-u)]^2} + (x_i \leftrightarrow x_j)\,.
\end{equation}
This result can be straightforwardly adapted to our set-up by rewriting the $d_\mu$ interactions as Yukawa couplings
(see eq.~(\ref{eq:d_mu_rewriting}))
\begin{equation}
\frac{1}{f} \partial_\mu h \sum_{i,j} c_{ij\textsc{l,r}} \overline \chi_i \gamma^\mu P_{L,R} \chi_j
\rightarrow \frac{1}{f} h\sum_{i,j} i c_{ij\textsc{l,r}} m_j \overline \chi_{iL,R} \chi_{jR,L} + {\rm h.c.}\,,
\end{equation}
corresponding to the following contributions to $y_{ij}$ and $\widetilde y_{ij}$
\begin{equation}
\Delta y_{ij} = i \frac{m_i - m_j}{\sqrt{2} f} c_{ij\textsc{l,r}} \,,
\qquad \quad
\Delta \widetilde y_{ij} = \mp \frac{m_i + m_j}{\sqrt{2} f} c_{ij\textsc{l,r}} \,.
\end{equation}
This formula shows that, if $d_\mu$ operators involving only left- or right-handed fermions are present,
$\Delta y_{ij}$ and $\Delta \widetilde y_{ij}$ always have the same complex phase. In this case,
the product of two $d_\mu$-symbol vertices $\Delta y_{ij} \Delta \widetilde y_{ij}^*$ appearing in eq.~(\ref{eq:w_diagram})
is real and does not lead to CP-violating effects. This explicitly proves that diagrams with two $d_\mu$ interactions do not contribute to the
Weinberg operator in the $\bf 14 +1$ set-up we are considering in this section.

The contribution to the Weinberg operator in eq.~(\ref{eq:w_diagram}) can be conveniently rewritten
by using a simple approximation for the $f_3$ function. If $x_{i,j} \gg 1$ the $f_3(x_i, x_j)$ function
is well approximated by $f_3 \simeq 1 - 1/3\bar x$, where $\bar x$ is the largest between $x_i$ and $x_j$.
For practical purposes, if one of the resonances in the loop has a mass $m \gtrsim 500\;$GeV, one can safely use
the approximation $f_3 = 1$. The only case in which this estimate is not fully accurate is for loops involving only
the top quark, in which case $f_3(x_{t}, x_{t}) \simeq 0.88$. Also in this case, however, the approximation $f_3 = 1$
is valid up to $\sim 10\%$ deviations.

By using straightforward algebraic manipulations, it can be shown that
\begin{eqnarray}
w &\simeq& \frac{g_s^3}{4 (4\pi)^4} \sum_{i,j} \frac{{\rm Re} [y_{ij} \widetilde y_{ij}^*]}{m_i m_j}\nonumber\\
&=& -\frac{g_s^3}{4 (4\pi)^4} {\rm Re\,Tr}\left[\frac{2}{f} \Upsilon \left(c_\textsc{r} M^{-1} - M^{-1} c_\textsc{l}\right)
- \frac{2}{f^2} i\, c_\textsc{r} M^{-1} c_\textsc{l} M + i\, \Upsilon M^{-1} \Upsilon M^{-1}\right]\,,\label{eq:w_app}
\end{eqnarray}
where $\Upsilon_{ij}$ denotes the matrix of Yukawa couplings, defined as
\begin{equation}
\sum_{i,j} h \Upsilon_{ij} \overline \chi_{iL} \chi_{jR} + {\rm h.c.}\,,
\end{equation}
and $M$ is the fermion mass matrix, defined as $\sum_{ij} M_{ij} \overline \chi_{iL} \chi_{jR} + {\rm h.c.}$.

\subsubsection{Neutron and Mercury EDM}

The quark electric and chromoelectric dipole operators and the Weinberg operator generate contributions
to the neutron EDM $d_n$.\footnote{Additional contributions to the neutron EDM can be generated by a top dipole moment
through running effects. If the top dipole is generated at loop level, as expected in many CH scenarios,
these corrections are however quite small and well below the current experimental bounds~\cite{Panico:2015jxa}.}
The explicit expression is given by~\cite{Brod:2013cka}
\begin{equation}\label{eq:nEDM}
\frac{d_n}{e} \simeq (1.0 \pm 0.5) \left[0.63 \left(\frac{d_d}{e} - 0.25\, \frac{d_u}{e}\right)
+ 1.1 \left(\widetilde d_d + 0.5\, \widetilde d_u\right) + 10^{-2}\,{\rm GeV}\, w\right]\,,
\end{equation}
where we took into account running effects from the top mass scale to the typical hadronic scale
$\mu_H \simeq 1\,{\rm GeV}$.\footnote{For simplicity we neglected additional running between the resonances
masses and the top mass.}

The CEDM's of the light quarks give also rise to EDM's for the diamagnetic atoms. At present the most stringent experimental
constraints come from the limits on the EDM of mercury (Hg). The latter can be estimated
as~\cite{Brod:2013cka}
\begin{equation}\label{eq:HgEDM}
\frac{d_{\rm Hg}}{e} \simeq - 0.9\cdot 10^{-4} \left(4^{+8}_{-2}\right)
\left(\widetilde d_u - \widetilde d_d - 0.76 \cdot 10^{-3}\,{\rm GeV}\, w\right)\,.
\end{equation}

It is interesting to compare the size of the various contributions to the neutron and mercury EDM's.
From eqs.~(\ref{eq:dq}) and (\ref{eq:dqtilde}) we can see that
\begin{equation}
d_q = \frac{8}{3} e Q_q \frac{\alpha}{\alpha_s} \widetilde d_q\simeq 0.06\, Q_q\, \widetilde d_q\,,
\end{equation}
where we set $Q_{f_i} = 2/3$, as in the model we consider in this section. The contributions to
$d_n$ coming from light quark EDM's is therefore suppressed by almost one order of magnitude
with respect to the one from the quark CEDM's.

Let us now consider the contributions from the Weinberg operator. Due to the different structure of the top partner contributions,
the effects due to the Weinberg operator and the ones from the Barr--Zee diagrams can not be exactly compared as we did for the
electric and chromoelectric moments. To get an idea of the relative importance we can however use a rough approximation,
namely
\begin{equation}\label{eq:w_est}
w \sim \frac{g_s^3}{(4\pi^4)} \frac{1}{f^2} {\rm Im}\,c_t \sim \frac{g_s}{4 m_q} \left(\frac{m_T}{f}\right)^2 \frac{1}{\log{m_T/m_t}} \widetilde d_q
\sim 40\,{\rm GeV}^{-1}\,\left(\frac{m_T}{f}\right)^2 \frac{1}{\log{m_T/m_t}} \widetilde d_q\,.
\end{equation}
This estimate is quite close to the exact result (eq.~(\ref{eq:w_14+1})), as we will see in sec.~\ref{sec:exp_bounds}.
An interesting feature of the contributions to the Weinberg operator is the fact that they are controlled by the compositeness scale $f$,
and are nearly independent of the top partner masses. As a consequence their relative importance with respect to the quark dipole
contributions grows for large $m_T/f$.

Using the estimate in eq.~(\ref{eq:w_est}) we find that, for $m_T \sim f$,
the $w$ contributions to the mercury EDM are suppressed by almost two orders of magnitude with respect to the quark CEDM's ones.
We thus expect the Weinberg operator to play a role for $d_{\rm Hg}$ only for sizable values of the ratio $m_T/f$, namely
$m_T/f \gtrsim 10$. On general ground one expects $m_T \sim g_* f$, with $g_*$ the typical composite sector coupling.
The contributions from the Weinberg operator to $d_{\rm Hg}$ are thus relevant only for new dynamics that are close to be
fully strongly-coupled.

The situation is significantly different for the neutron EDM. In this case the contributions from the Weinberg operator are suppressed
by a factor $\sim 1/4$ if the top partners are light ($m_T/f = 1$). For heavier partner masses, $m_T/f \gtrsim 3$, the bounds coming from
the Weinberg operator can thus become competitive with the ones from the quark CEDM's. We will discuss this point more quantitatively
in the following.

\subsection{Experimental bounds}\label{sec:exp_bounds}

We can now discuss the constraints coming from the experimental data. The present searches for electron~\cite{Baron:2013eja}, neutron~\cite{Baker:2006ts} and mercury~\cite{Graner:2016ses} EDM's give null results and can thus be used to extract the following constraints
\begin{eqnarray}
\left|d_e\right| &< 9.4\cdot 10^{-29}\,e\,\text{cm}\qquad & {\rm at}\ \ 90\%\; {\rm CL}\,,\\
\rule{0pt}{1.25em}\left|d_n\right| &< 2.9\cdot 10^{-26}\,e\,\text{cm}\qquad & {\rm at}\ \ 95\%\; {\rm CL}\,,\\
\rule{0pt}{1.25em}\left|d_{\rm Hg}\right| &< 7.4\cdot 10^{-30}\,e\,\text{cm}\qquad & {\rm at}\ \ 95\%\; {\rm CL}\,.
\end{eqnarray}
Near-future experiments are expected to significantly improve the bounds on the neutron and electron EDM's.
The neutron EDM bounds could be improved up to $|d_n| < 10^{-27}\,e\,\text{cm}$~\cite{Picker:2016ygp}. On the other hand, the ACME collaboration estimates the future sensitivity on the electron EDM to be~\cite{Doyle}
\begin{equation}
|d_e| \lesssim 0.5\cdot 10^{-29}\,e\,{\rm cm}\qquad (\text{ACME II})
\end{equation}
and
\begin{equation}
|d_e|\lesssim 0.3\cdot 10^{-30}\,e\,{\rm cm}\qquad (\text{ACME III})
\end{equation}
that correspond to an improvement of the current constraints by more than two orders of magnitude.\footnote{An additional
bound on the electron EDM has been reported in ref.~\cite{Cairncross:2017fip}, $\left|d_e\right| < 1.3\cdot 10^{-28}\,e\,\text{cm}$ at $90\%\;$CL, which is slightly weaker than the current ACME constraint. This experiment
is currently limited by statistics and in the future is expected to allow for a precision $\sim 10^{-30}\,e\,\text{cm}$.}

It is interesting to compare the impact of the different bounds on the parameter space of composite Higgs models.
An easy way to perform the comparison is to focus on the constraints on the EDM of the electron and on the EDM's
and CEDM's of the light quarks. As can be seen from eqs.~(\ref{eq:e-EDM}), (\ref{eq:dq}) and (\ref{eq:dqtilde})
in the $\bf 14 +1$ model with a light fourplet all these effects depend on the quantity\footnote{As we discussed before, in the
$\bf 14 + 1$ with a light fourplet only charge-$2/3$ partners contribute to Barr--Zee diagrams, thus $Q_{f_i} = 2/3$ in eqs.~(\ref{eq:e-EDM}) and (\ref{eq:dq}).}
\begin{equation}
\widetilde \gamma \equiv \frac{v}{f} \sum_i (c_{i\textsc{r}} - c_{i\textsc{l}}) f_1(x_i)\,.
\end{equation}
The bounds on $\widetilde \gamma$ can thus be used to compare the strength of the various experimental searches.
For simplicity we will neglect corrections coming from the Weinberg operator, and we will assume that the
electron and light quark Yukawa's coincide with the SM ones.

The constraints from the electron EDM measurements read
\begin{equation}
\begin{array}{l@{\qquad}l}
\rule{0pt}{.25em}\left|\widetilde \gamma\right| < 0.029\quad & {\rm current\ bound}\,,\\
\rule{0pt}{1.25em}\left|\widetilde \gamma\right| \lesssim 1.5 \times 10^{-3}\quad & {\rm ACME\ II}\,,\\
\rule{0pt}{1.25em}\left|\widetilde \gamma\right| \lesssim 1.0 \times 10^{-4}\quad & {\rm ACME\ III}\,.
\end{array}
\end{equation}
The bounds from the neutron EDM measurement are
\begin{equation}
\begin{array}{l@{\qquad}l}
\rule{0pt}{.25em}\left|\widetilde \gamma\right| < [0.08, 0.23]\quad & {\rm current\ bound}\,,\\
\rule{0pt}{1.25em}\left|\widetilde \gamma\right| \lesssim [0.003, 0.01]\quad & {\rm improved\ bound}\,.
\end{array}
\end{equation}
Finally the bounds from the mercury EDM are
\begin{equation}
\left|\widetilde \gamma\right| < [0.06, 0.4]\,.
\end{equation}
Notice that for the neutron and mercury EDM bounds we took into account the error range in the
estimates in eqs.~(\ref{eq:nEDM}) and (\ref{eq:HgEDM}).

From the above results we find that, at present, the electron EDM measurements give the strongest constraints.
The future improvements on the neutron EDM constraints could strengthen the present electron EDM bounds by a
factor of order $3$. These constraints, however, will be easily surpassed by the new electron EDM experiments,
which can improve the current bounds by a factor of $\sim 20$ in the near future (ACME II) and by more than
two orders of magnitude afterwards (ACME III).

\begin{figure}
\centering
\includegraphics[width=.485\textwidth]{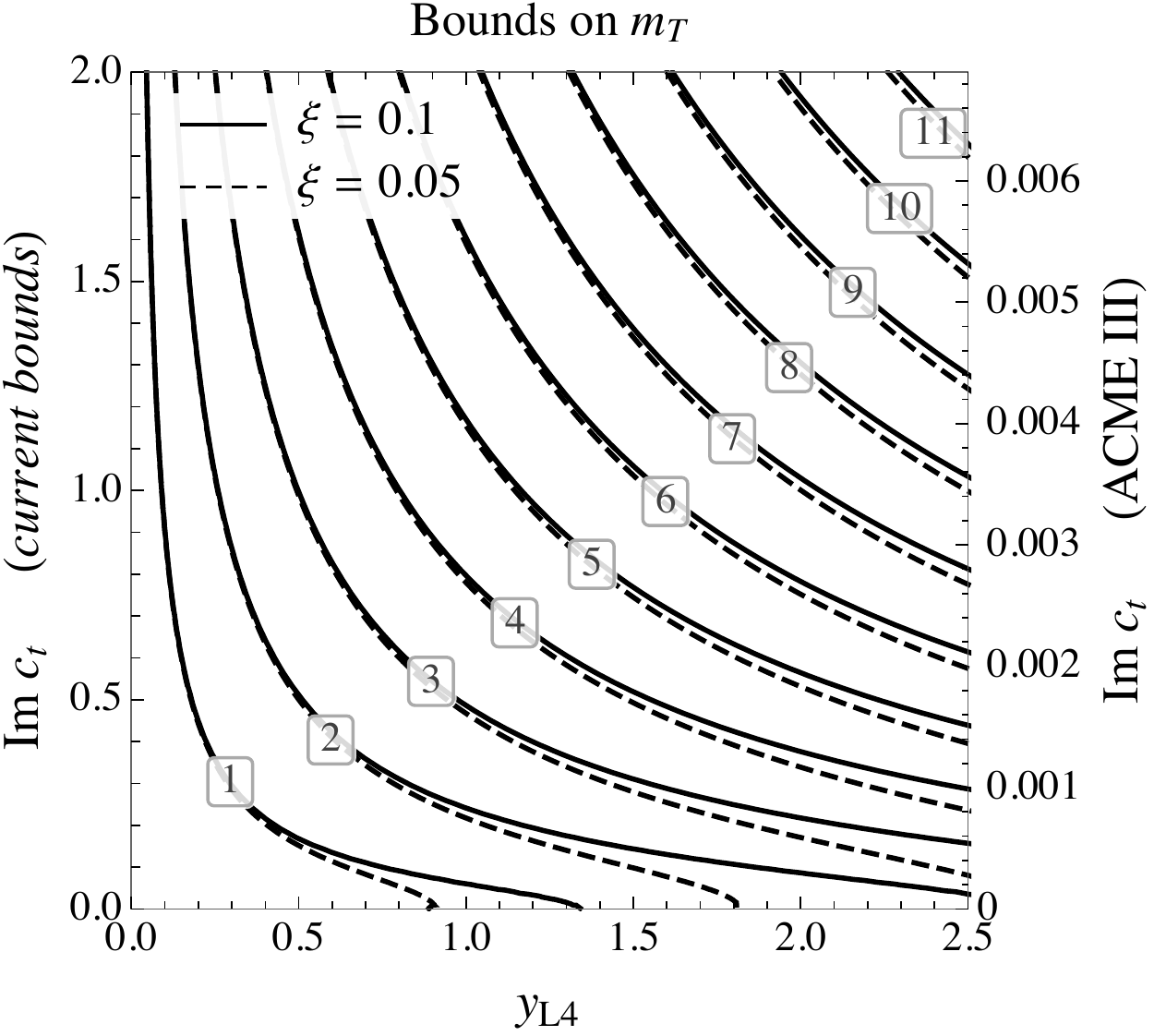}
\hfill
\includegraphics[width=.485\textwidth]{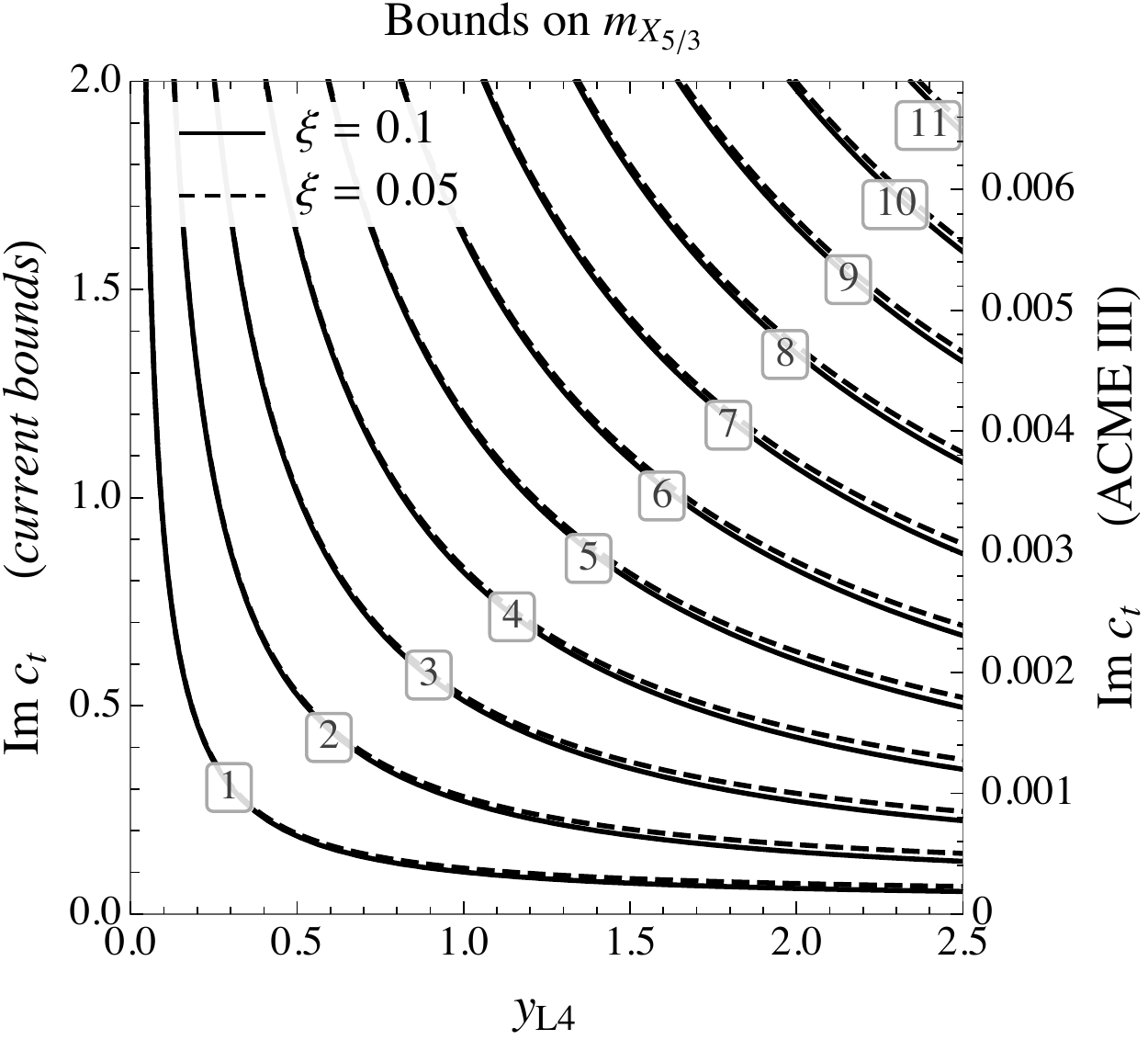}
\caption{Bounds on the mass of the $T$ (left panel) and $X_{5/3}$ (right panel) states derived from the constraints on the electron EDM. The bounds are expressed in TeV  and are presented as a function of the elementary--composite mixing $y_{L4}$ and of the imaginary part of $c_t$. The labels on the left vertical axis corresponds to the present
bounds, while the ones on the right axis correspond to the ACME III projections.
The solid and dashed lines correspond to the choice $\xi = 0.1$ and $\xi = 0.05$ respectively.
}\label{fig:e-EDM_bounds}
\end{figure}

The constraints on the top partner masses in the ${\bf 14 + 1}$ scenario are shown in fig.~\ref{fig:e-EDM_bounds} as a function
of the $y_{L4}$ mixing parameter and of the imaginary part of the $c_t$ coupling. The value of the $y_{Lt}$ mixing has been
fixed by requiring that the correct top mass is reproduced. In the left panel we show the bounds on the
mass of the $T$ partner, while in the right panel we show the bounds on the mass of the lightest top partner in the multiplet, namely
the $X_{5/3}$ state.
The solid and dashed lines show the bounds for $\xi \equiv v^2/f^2 = 0.1$ and $\xi = 0.05$ respectively, which roughly correspond
to the present constraints on $\xi$
coming from Higgs couplings measurements~\cite{Aad:2015pla} and to the projected bounds for high-luminosity LHC~\cite{fut_Higgs_couplings,Dawson:2013bba}.
The impact of
$\xi$ on the bounds is however quite mild. Notice that the $T$ mass, even without any constraint from the electron EDM
(i.e.~for ${\rm Im} c_t = 0$) is still bounded from below. This is due to the fact that, even setting $m_4 = m_{X_{5/3}} = 0$,
$m_T$ still gets a contribution  from the mixing with the elementary states, which translates into $m_T = |y_{L4} f|$.

Using simple power counting considerations~\cite{Giudice:2007fh,DeSimone:2012fs} we can estimate the typical size
of the $y_{L4}$ and $c_t$ parameters to be $y_{L4} \sim y_{Lt} \sim y_{top}$ and $c_t \sim 1$. Barring accidental suppressions
in the complex CP-violating phase of $c_t$, we get that the present constraints from the electron EDM correspond to bounds
on the top partner masses in the range $2 - 4\;$TeV. The ACME II experiment will extend the exclusion range to masses
of order $10 - 20\;$TeV, whereas masses in the range $50 - 100\;$TeV will be tested by ACME III.

Another useful way to quantify the strength of the electron EDM bounds is to fix the mass of the top partners and derive the
amount of suppression needed in the complex phase of $c_t$ to pass the experimental bounds. Choosing masses of order
$3\;$TeV, roughly of the order of the possible direct bounds from high-luminosity LHC, we can see that the present constraints
still allow for order one complex phases. ACME II will lower the bound to $\sim 5\%$, while ACME III will be able to constrain
CP-violating phases significantly below the $1\%$ level.

It is important to stress that the bounds coming from the electron and light quark EDM's crucially depend on the assumption that the
light fermion Yukawa couplings are not (strongly) modified with respect to the SM predictions. If the
light fermion masses are generated through partial compositeness, this assumption is typically satisfied. One indeed expects all
Yukawa couplings to deviate from their SM values only by corrections of order $\xi$. The current bounds $\xi \lesssim 0.1$
guarantee that the Yukawa couplings agree within $\sim 10\%$ with their SM values.

It is however conceivable that substantial modifications of the partial compositeness structure could exist for the light fermions.
In such a case large deviations of the Yukawa couplings could be present. Strong suppression in some or all the light fermion Yukawa's
would modify the relative importance of the constraints coming from the experimental measurements.
As we discussed before, the contributions to the electron EDM are controlled by the electron Yukawa, whereas the light quark EDM and
CEDM are proportional to the $u$ and $d$ Yukawa's. The experimental constraints on the electron and neutron EDM thus
carry complementary information and can become more or less relevant in different contexts.

It is interesting to notice that the contributions to the Weinberg operator are independent of the light fermion Yukawa's and only depend
on the top and top partners couplings to the Higgs. They can thus be used to extract bounds that are in principle more model independent
than the ones coming from the electron and light quark EDM's. Using the approximation in eq.~(\ref{eq:w_app}), we find that the
contribution to the Weinberg operator in the $\bf 14 + 1$ model with a light fourplet is
\begin{equation}\label{eq:w_14+1}
w \simeq - \frac{g_s^3}{2 (4 \pi)^4 f} {\rm Re\,Tr}[\Upsilon c_\textsc{r} M^{-1}]
= \frac{2 g_s^3}{(4\pi)^4}\frac{\sqrt{2}\, y_{L4}}{f^2 y_{Lt}} {\rm Im}\,c_t
\simeq \frac{2 g_s^3}{(4\pi)^4}\frac{y_{L4} m_4}{\sqrt{m_4^2 + y_{L4}^2 f^2}} \frac{v}{f^2\, m_{top}} {\rm Im}\,c_t\,.
\end{equation}
A noteworthy aspect of this formula is the fact that it depends on the top partners masses only indirectly. The dependence on $m_4$ only
appears when we rewrite the $y_{Lt}$ parameter as a function of the top mass. This feature indicates that the contributions to the
Weinberg operator are not controlled by the lightest resonances, as was the case for the dipole operators, but instead can receive sizable
contributions from the UV dynamics. Of course, since the IR and UV contributions are in general independent, we do not expect them
to cancel each other. The result in eq.~(\ref{eq:w_14+1}) can thus be used as a lower estimate to obtain constraints on the parameter
space of the model.

\begin{figure}
\centering
\includegraphics[width=.5\textwidth]{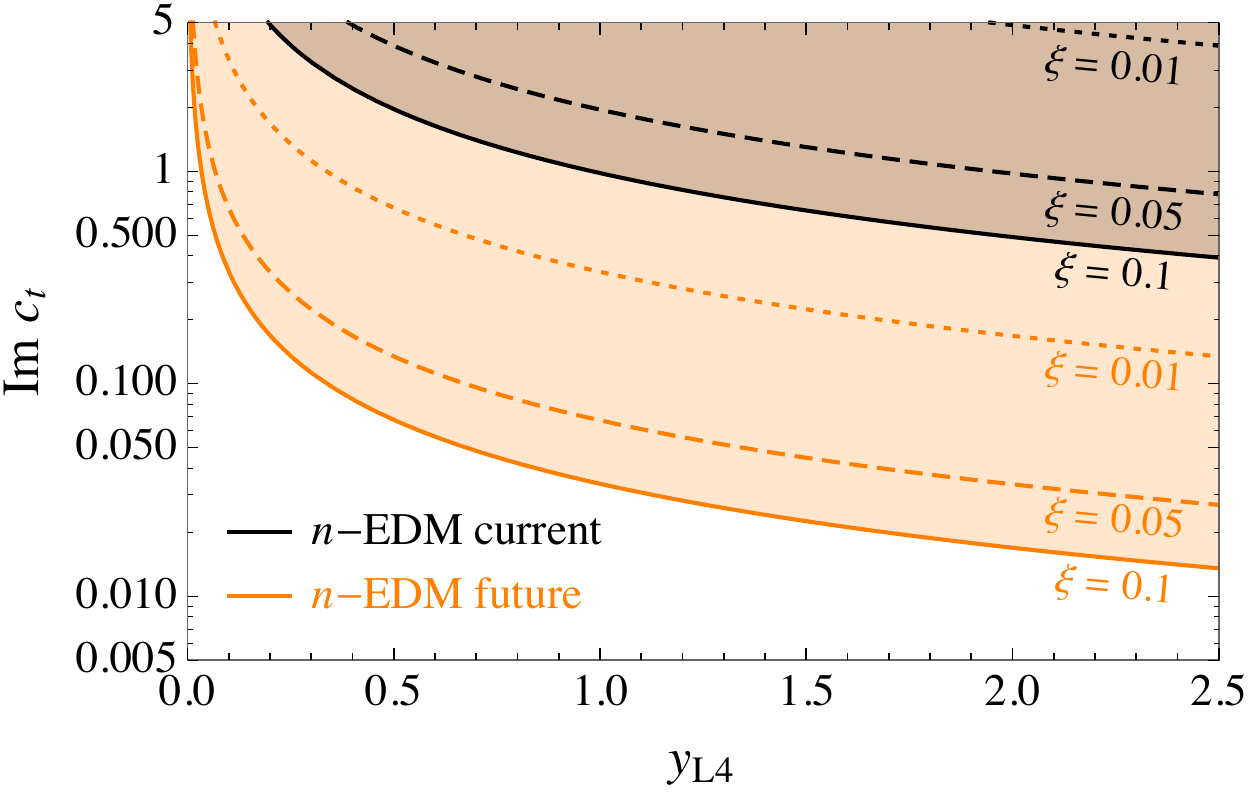}
\caption{Bounds on the CP-violating part of the $c_t$ coupling as a function of the $y_{L4}$ mixing derived from the current and projected
constraints on the neutron EDM. The results are derived by using the constraints on the Weinberg operator
}\label{fig:bounds_WeinbergOp}
\end{figure}
In fig.~\ref{fig:bounds_WeinbergOp} we show the bounds in the $({\rm Im}\,c_t,\, y_{L4})$ plane coming from the current (black lines)
and projected (orange lines) neutron EDM measurements for various values of $\xi$ ($\xi = 0.1, 0.05, 0.01$).
These results are obtained by taking into account only the contributions from the Weinberg
operator in eq.~(\ref{eq:nEDM}) (we use the lower estimate of the effect to derive the numerical results),
and neglecting the ones from the light-quark dipole operators. Notice that we also neglected additional contributions to
the Weinberg operator that can be induced by the presence of a top CEDM~\cite{Kamenik:2011dk}. These effects are of order
\begin{equation}
w_{t-\textsc{cedm}} = \frac{g_s^3}{32 \pi^2} \frac{\widetilde d_t}{m_t} \simeq \frac{g_s^3}{32 \pi^2} \frac{1}{16 \pi^2 f^2}\,.
\end{equation}
and are subleading with respect to the contributions in eq.~(\ref{eq:w_14+1}) if $y_{L4}\, \rm{Im}\,c_t \gtrsim 0.2$. As can be seen from
fig.~\ref{fig:bounds_WeinbergOp} these effects are irrelevant for the present constraints. They are instead expected to become comparable
with the top partners contributions in part of the parameter space probed by future experiments. In this situation the constraint
given in fig.~\ref{fig:bounds_WeinbergOp} can still be considered as a lower bound, provided strong accidental cancellations do not occur.

We can see that, for $\xi = 0.1$, the current neutron EDM constraints
typically forbid values of ${\rm Im}\;c_t$ larger than $\sim 1$. These bounds are competitive with the current ones from the
electron EDM (see fig.~\ref{fig:e-EDM_bounds}) if the top partner masses are $m_{X_{5/3}} \gtrsim 5 - 6\;$TeV, whereas they are
weaker for lighter resonances. Notice that the bound from the Weinberg operator roughly scales like $f^{-2}$, so it quickly degrades for
smaller values of $\xi$. The bound from the electron EDM has instead a much milder dependence on $\xi$.

Future improvements on the neutron EDM measurements (orange lines in fig.~\ref{fig:bounds_WeinbergOp}) could strengthen the bounds
by more than one order of magnitude. The improved bounds, for $\xi = 0.1$, would be comparable to the present ones
from the electron EDM for $m_{X_{5/3}} \simeq 1\;$TeV. Notice however that the projected improvement in the electron EDM constraints
(ACME III) would make the Weinberg operator bounds relevant only for very heavy top partners ($M_{X_{5/3}} \gtrsim 20\;$TeV).

\medskip

As we mentioned in the Introduction, the bounds we presented in this section apply directly to models in which the
flavor structure is implemented through a ``dynamical scale'' mechanism (see ref.~\cite{Panico:2016ull}). In these scenarios
direct CP-violating effects involving the light SM fermions are strongly suppressed and the leading effects are generated
only from two-loop contributions involving the top and its partners. In other flavor scenarios, for instance 
anarchic partial compositeness models, additional sizable CP-violating contributions can be present. We will briefly discuss
these effects in the following.

In anarchic partial compositeness models, corrections to the light quark EDM's and CEDM's are typically generated at one loop~\cite{Konig:2014iqa} (see ref.~\cite{Panico:2015jxa} for a review). For a quark $q$ these effects can be estimated as
\begin{equation}
\frac{d_q}{e} \sim \widetilde d_q \sim \frac{m_q}{16 \pi^2}\frac{1}{f^2}\,.
\end{equation}
These contributions are roughly one inverse loop factor $16 \pi^2/g_s^2 \simeq 10^2$ larger than the Barr--Zee effects,
thus, barring accidental cancellations, are usually dominant.
The current neutron EDM constraints lead to a lower bound $f \gtrsim 4.5\;\rm{TeV}$ coming from the down-quark dipole operator.
A slightly weaker constraint, $f \gtrsim 2\;\rm{TeV}$, is obtained from the up-quark dipole.

If the anarchic structure is naively extended to the lepton sector, large one-loop contributions to the electron EDM are present.
The current bounds on the electron EDM imply a  constraint $f \gtrsim 38\;\rm{TeV}$, which rules out top partners in the
$50 -100\;\rm{TeV}$ range. In these scenarios a similar bound also comes from the lepton flavor violating
decay $\mu \rightarrow e \gamma$.

We finally consider models with flavor symmetries. In the case of $\U(3)$ symmetry~\cite{Redi:2011zi},
the one-loop contributions to the light-quark EDM's are comparable to the ones in anarchic scenarios. A significant suppression
of these effects can instead be present in $\U(2)$ models~\cite{Barbieri:2012uh} if the partners of the light quarks are decoupled.
In this case the two-loop Barr--Zee contributions become dominant and the bounds derived in this section apply.

\subsection{Comparison with direct top partner searches}

It is also interesting to compare the bounds from CP-violating effects with the direct searches for top partners.
We start the discussion by considering the constraints coming from the LHC. The strongest bounds on the mass of a light fourplet
come from searches for the exotic charge-$5/3$ top partner, the $X_{5/3}$, which decays exclusively into $Wt$.
So far the experimental searches focussed
mainly on top partners pair production. The strongest bounds come from searches
in the lepton plus jets final state, whose present constraints are $m_{X_{5/3}} > 1250\;$GeV (ATLAS collaboration~\cite{Aaboud:2017zfn}) and $m_{X_{5/3}} > 1320\;$GeV (CMS collaboration~\cite{CMS:2017wwc}).

Additional bounds come from searches in the same-sign dilepton final state, whose sensitivity is only slightly lower than
the one in the lepton plus jets channel. The present bounds for pair-produced top partners are $m_{X_{5/3}} > 1160\;$GeV from the
CMS analysis in ref.~\cite{CMS:2017jfv} and $m_{X_{5/3}} > 990\;$GeV from the ATLAS analysis
in ref.~\cite{ATLAS:2016sno}.\footnote{The ATLAS analysis is only available for $3.2/$fb integrated luminosity at $13\;$TeV.
This explains the significantly lower bound with respect to the CMS analysis, which instead exploits $35.9/$fb integrated luminosity.}
Interestingly, searches for charge-$5/3$ resonances in same-sign dileptons are sensitive not only to pair production
but also to single production. This aspect was investigated in ref.~\cite{Matsedonskyi:2014mna} for the $8\;$TeV LHC searches.
The same-sign dilepton search was found to be sensitive to single production with relatively high efficiencies,
namely $\sim 50\%$ of the pair-production signal efficiency for the ATLAS search and $\sim 10\%$ for the CMS one.
The $13\;$TeV searches are analogous to the $8\;$TeV ones, so one expects similar efficiencies to apply.
The sensitivity to single production can significantly enhance the bounds for large values of $c_t$.
Indeed this coupling controls the $W X_{5/3} t$ vertex~\cite{Matsedonskyi:2015dns},
\begin{equation}\label{eq:gW_coupl}
g_{W X_{5/3} t_R} = \frac{g}{\sqrt{2}}c_t\frac{v}{f}\,,
\end{equation}
that mediates single production in association with a top quark.\footnote{Experimental searches for
singly-produced heavy quarks decaying into $Z\,t/b$~\cite{Sirunyan:2017ynj}, $h\,t/b$~\cite{Khachatryan:2016vph} and $Wb$~\cite{ ATLAS:2016ovj, Sirunyan:2017tfc}
are also available in the literature. The bounds from these
searches on fourplet top partners are however weaker than the ones we derived with the recast of the same-sign
dilepton searches.}

Interestingly, the searches in lepton plus jets and same-sign dilepton final states are sensitive
not only to charge-$5/3$ resonances but also to states with charge $-1/3$ decaying into $Wt$.
The bounds reported in the experimental analyses for resonances with charge $5/3$ and $-1/3$ are quite close,
thus signaling similar search efficiencies. A reasonable estimate of the bounds can thus be obtained by just adding
the production cross sections for both types of partners.
As we discussed before, the fourplet multiplet contains a state with charge $-1/3$, the $B$, which decays into $Wt$
with a branching ratio close to $100\%$.
If the mass split between the $X_{5/3}$ and $B$ states is below $\sim 200\;$GeV, which requires relatively small
value of $y_{L4}$ ($y_{L4} \lesssim 1$ in the case $m_{X_{5/3}} \sim 1 - 2\;{\rm TeV}$ and $\xi \simeq 0.1$),
the same-sign dilepton signal is enhanced by almost a factor $2$, with a significant impact on the
exclusion bounds~\cite{Matsedonskyi:2014mna,Matsedonskyi:2015dns}.

\begin{figure}
\centering
\includegraphics[width=.485\textwidth]{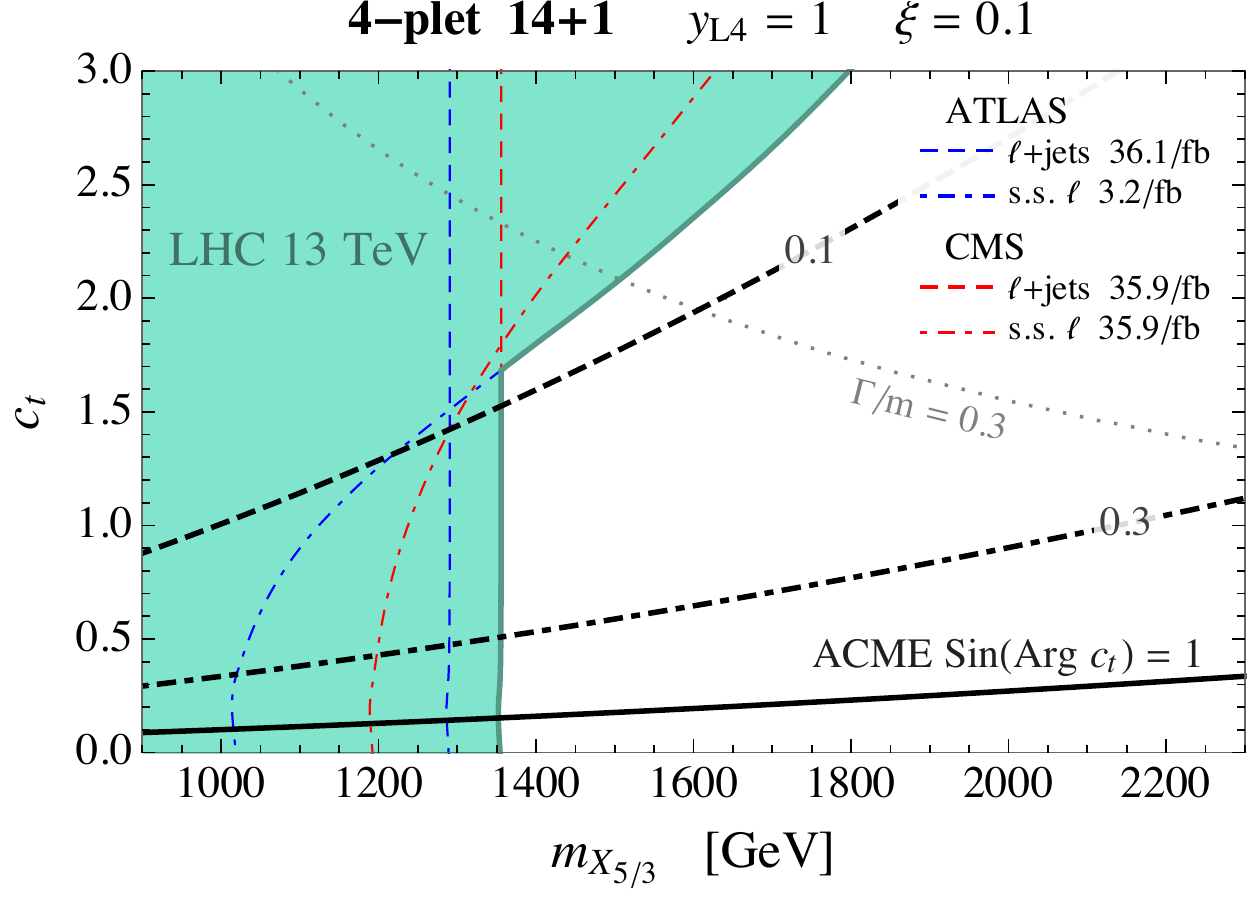}
\hfill
\includegraphics[width=.485\textwidth]{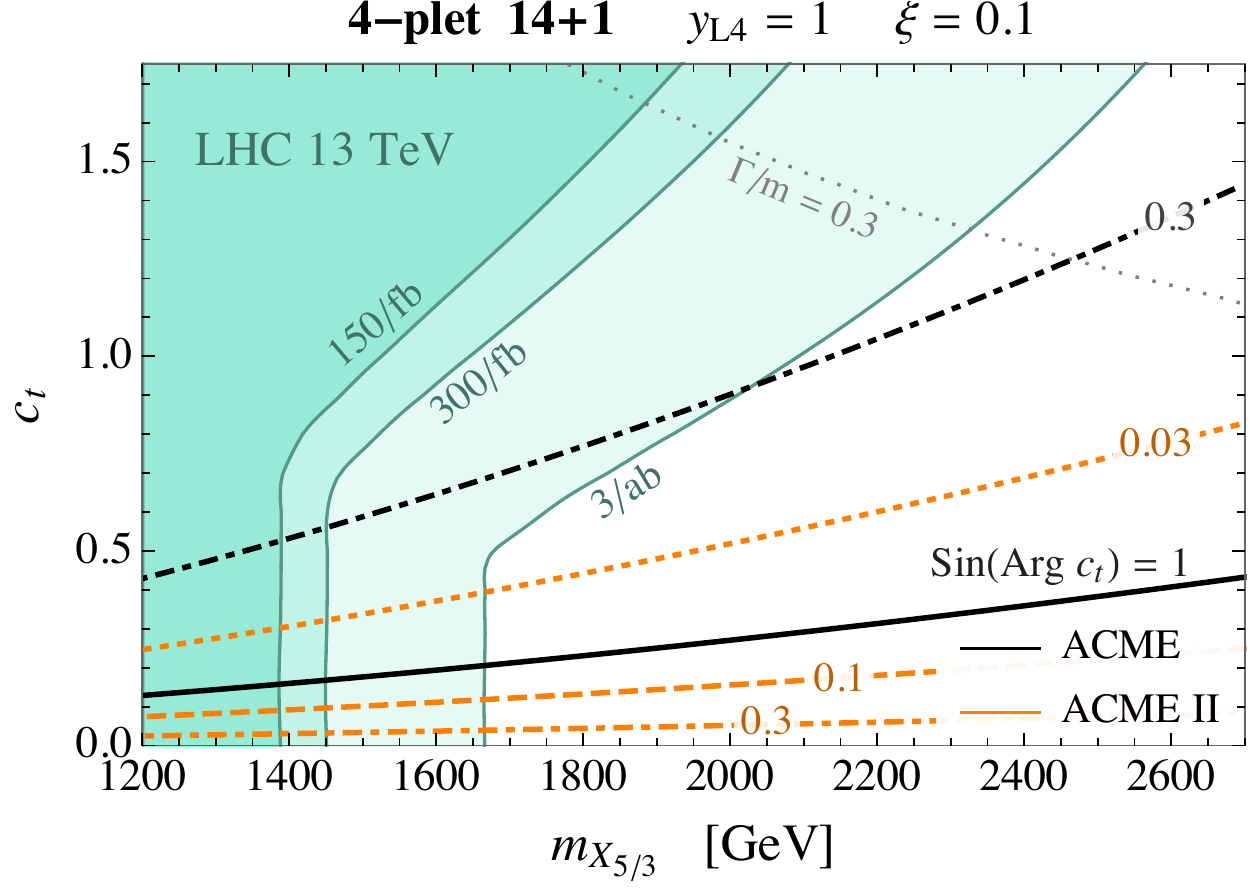}
\caption{Bounds on the $c_t$ coupling as a function of the mass of the $X_{5/3}$ resonance for the scenario with a light
fourplet in the $\bf 14 +1$ model (for the choice $\xi = 0.1$ and $y_{L4} = 1$).
The current bounds from the LHC data and from the constraints on the electron EDM are shown in the left panel, whereas the projections for the future LHC runs and the estimate of the future ACME II constraints are shown in the right panel.
In the left panel we also show separately the direct bounds from the lepton plus jets (dashed lines) and for the same-sign dilepton
analyses (dot-dashed lines) for ATLAS (blue) and CMS (red). The bound from the electron EDM current (black lines) and improved ACME II searches
(orange lines) are shown for different choices of the complex phase of $c_t$ ($\sin({\rm Arg}\, c_t) = 1, 0.3, 0.1, 0.03$ for the solid, dashed, dot-dashed and dotted lines respectively). In the region above the dotted gray line the width of the $X_{5/3}$ resonance is
above $30\%$ of its mass. 
}\label{fig:comparison_with_direct}
\end{figure}
The direct bounds on the mass of the $X_{5/3}$ resonance from the LHC searches are shown by the shaded green regions
in fig.~\ref{fig:comparison_with_direct}. The current bounds are shown in the left panel, while the projections for the future LHC runs
are in the right panel. For definiteness we set $\xi = 0.1$ (which roughly corresponds on the bound coming from
precision electroweak tests~\cite{Grojean:2013qca} and from present Higgs couplings measurements~\cite{Aad:2015pla})
and $y_{L4} = 1$. We also fix $y_{R4}$ by requiring the top mass to have the correct value.

As we discussed before, the strongest indirect constraints from CP-violating effects come from the electron EDM measurements.
The current bounds are shown in the figure by the black lines, while the ACME II projections are given by the orange lines. The bounds
are presented for different values of the complex phase of $c_t$, namely $\sin({\rm Arg}\, c_t) = 1, 0.3, 0.1, 0.03$.
One can see that indirect bounds tend to be stronger than the ones from direct searches for larger values of the top partners masses.
If the complex phase of $c_t$ is not too small, $\sin({\rm Arg}\, c_t) \gtrsim 0.1$, the current ACME constraints can easily probe
resonance masses $\sim 2\;$TeV, which are not tested by the run-2 LHC data. Moreover it can be seen that the additional parameter space
region probed by taking into account single production (corresponding to the improved LHC bounds at large $c_t$) can be also covered
by the electron EDM constraints if $\sin({\rm Arg}\, c_t) \gtrsim 0.1$ for current searches and $\sin({\rm Arg}\, c_t) \gtrsim 0.05$
for the high-luminosity LHC and ACME II.

For different values of $\xi$ the results in fig.~\ref{fig:comparison_with_direct} change only mildly. The indirect bounds are nearly
unaffected, while the direct searches are modified due to the rescaling of the single production coupling
(see eq.~(\ref{eq:gW_coupl})).
The dependence of the direct bounds on $y_{L4}$ is also mild, since this parameter only controls the split between the $X_{5/3}$ and
$B$ masses. The bound on $c_t$ coming from the electron EDM instead scales roughly linearly with $y_{L4}$ as can be seen
from eqs.~(\ref{eq:cT_value}) and (\ref{eq:eEDM_app}).

\begin{figure}
\centering
\includegraphics[width=.485\textwidth]{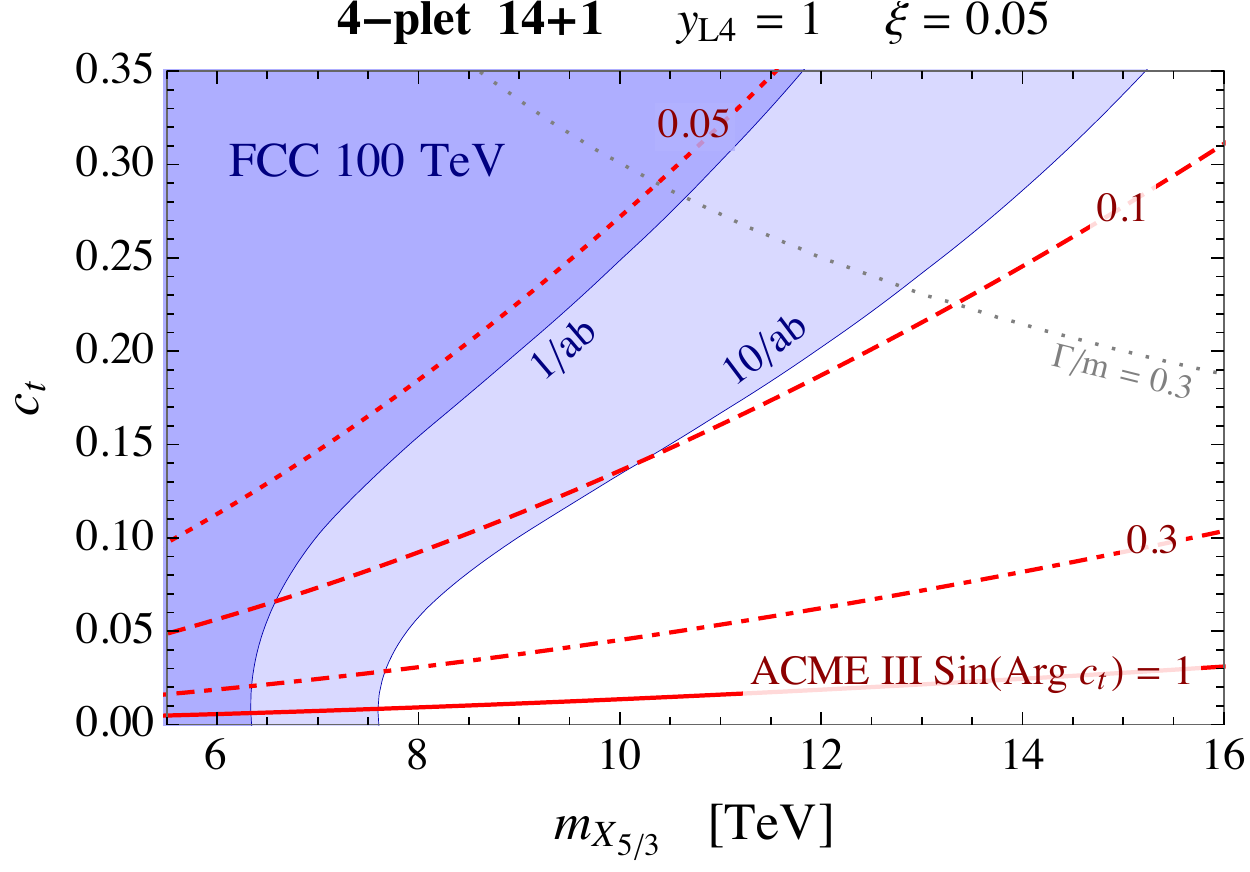}
\hfill
\includegraphics[width=.485\textwidth]{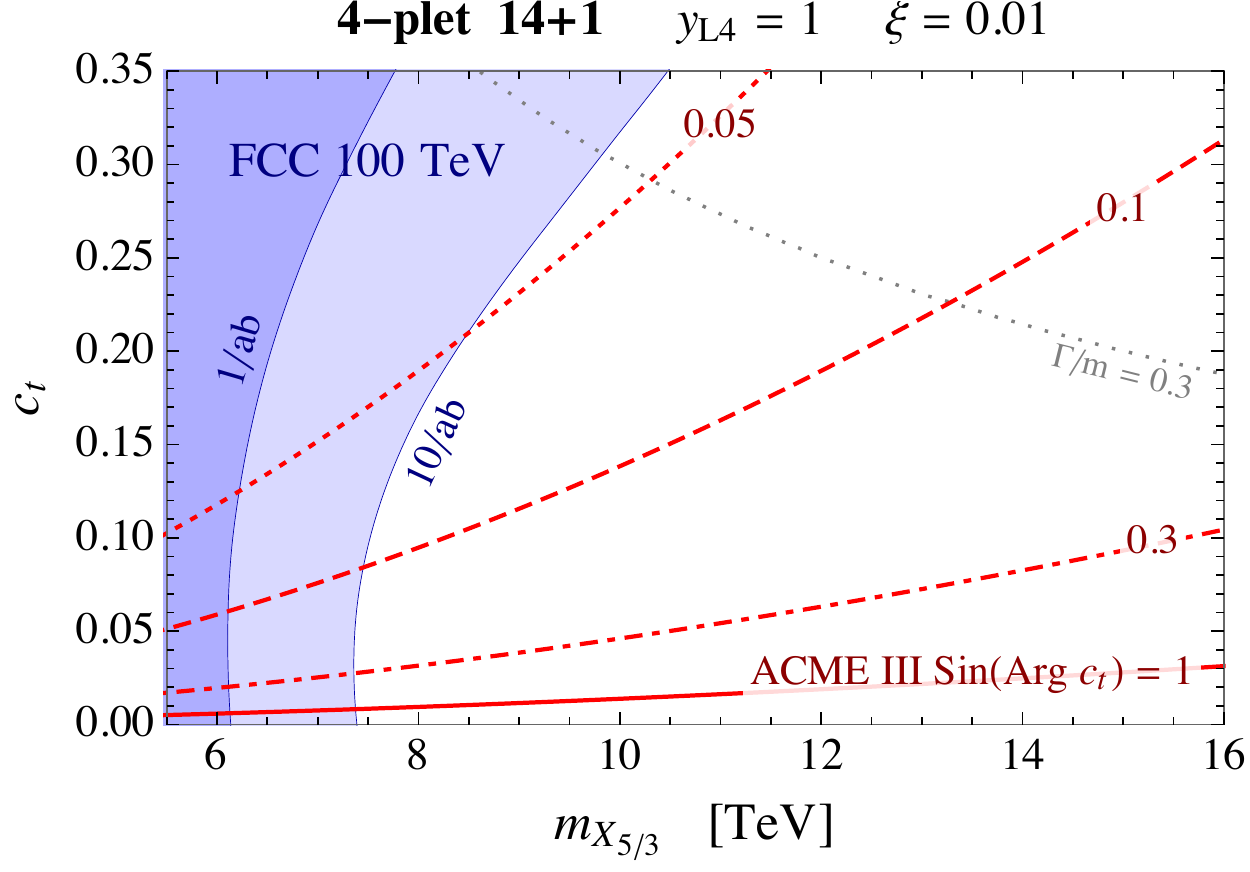}
\caption{Future direct and indirect exclusion bounds on the $c_t$ coupling as a function of the mass of the $X_{5/3}$ resonance
for the scenario with a light fourplet in the $\bf 14 +1$ model (for the choice $y_{L4} = 1$). The left and right panels correspond
to $\xi = 0.05$ and $\xi = 0.01$ respectively. The direct bounds from top partners searches at FCC-hh are given by the blue shaded regions
(for integrated luminosities $1/$ab and $10/$ab). The red lines correspond to the indirect exclusions for the estimated ACME III
sensitivity.
}\label{fig:comparison_with_direct_fcc}
\end{figure}
Finally, in fig.~\ref{fig:comparison_with_direct_fcc}, we compare the estimate for the direct exclusion reach at a future $100\;$TeV hadron
machine (FCC-hh) with the indirect bounds from the estimates of the ACME III sensitivity. In the left panel we set $\xi = 0.05$ which roughly
corresponds to the high-luminosity LHC reach, while in the right panel we set $\xi = 0.01$ which is the projected sensitivity at a
high-energy linear lepton collider (eg.~ILC at $500\;$GeV center of mass energy with $\sim 500/$fb integrated luminosity~\cite{Dawson:2013bba}). As one can see, in the absence of strong suppressions in the complex phase of $c_t$,
the ACME III reach can easily surpass the FCC-hh ones in a large part of the parameter space of the $\bf 14 + 1$ model.

\section{Non-minimal models}\label{sec_non_minimal_models}

In order to highlight the main features of CP-violation due to the top partners, in the previous section we focussed
on a simplified scenario with only one light multiplet. In generic realizations of the composite Higgs idea, however,
it is not uncommon to find non-minimal set-ups with multiple light top partners. In the following we will discuss
how the results we got in the simplified $\bf 14 +1$ model are modified in the presence of additional light resonances.
In addition we will consider an alternative scenario in which both the left-handed and right-handed top quark
components are realized as elementary states. This set-up can be interpreted as an effective description of the
MCHM$_5$ holographic scenario~\cite{Agashe:2004rs}.

\subsection{The $\bf 14 + 1$ model with a light singlet}

As a first example we consider a more complete version of the $\bf 14 + 1$ model, including not only a light fourplet,
but also a light singlet.
The Lagrangian of the model is given by the terms in eq.~(\ref{eq:Lcomp_14+1}) plus the following additional operators involving
the singlet $\psi_1$
\begin{eqnarray}
{\cal L} &=& i \overline \psi_1 \slashed D \psi_1 - \left(m_1 \overline \psi_{1L} \psi_{1R} + {\rm h.c.}\right)\nonumber\\
&& + \left( y_{L1} f (U^t \overline q_L^{\bf 14} U)_{55} \psi_{1R} - i c_L \overline \psi_{4L}^i \gamma^\mu d_\mu^i \psi_{1L}
- i c_R \overline \psi_{4R}^i \gamma^\mu d_\mu^i \psi_{1R} + {\rm h.c.}\right)\,.
\end{eqnarray}

The above Lagrangian contains four free parameters, that are in general complex. By field redefinitions two
parameters can be made real, thus leaving two additional CP-violating sources corresponding to the complex
phases of the combinations $c_L m_1 m_4^* y_{L1}^* y_{L4}$ and $c_R y_{L1}^* y_{L4}$. A convenient choice of phases
is obtained by making the mass parameter $m_1$ and the elementary-composite mixing $y_{L1}$ real. This choice
makes manifest that CP-violating effects are necessarily related to the $d_\mu$-symbol operators,
and are controlled by the $c_L$ and $c_R$ parameters (on top of the $c_t$ parameter we discussed in the previous section).

The mass of the singlet eigenstate $\widetilde T$ is
\begin{equation}
m_{\widetilde T} \simeq |m_1|\left[1 + \frac{1}{4}\frac{y_{L1}^2 f^2}{m_1^2}\frac{v^2}{f^2} + \cdots\right]\,.
\end{equation}
while the spectrum of the remaining states coincides with the one described in section~\ref{sec:e-EDM},
apart from modifications arising at higher order in $v/f$.

The CP-violating Higgs couplings to the top partners are given by
\begin{equation}\label{eq:d_mu-couplings_14+1_full}
-i\, c_{L,R} \overline \psi_{4L,R}^i \gamma^\mu d_\mu^i \psi_{1L,R} + {\rm h.c.} \supset i \frac{c_{L,R}}{f} \partial_\mu h
\left(\overline {\widehat X}_{2/3L,R} \gamma^\mu \widetilde T_{L,R} - \overline {\widehat T}_{L,R} \gamma^\mu \widetilde T_{L,R}\right) + {\rm h.c.}\,,
\end{equation}
where we only included the leading order terms in the $v/f$ expansion. As in the simplified set-up we discussed in
the previous section, also in the extended $\bf 14+1$ model the CP-violating effects arise only from charge $2/3$ fields.

In the mass-eigenstate basis the coefficients of the CP-violating interactions that give rise to Barr--Zee-type
contributions (see eq.~(\ref{eq:HGG_matrix_element})) read
\begin{equation}
\left\{
\begin{array}{l}
\displaystyle c_{top,L} = \sqrt{2} v \frac{y_{L1} y_{L4} m_4 f}{m_1(m_4^2 + y_{L4}^2 f^2)} {\rm Im}\, c_L\\
\rule{0pt}{1.75em}\displaystyle c_{T,L} = \sqrt{2} v \frac{y_{L1} y_{L4} m_1 m_4 f}{(m_4^2 + y_{L4}^2 f^2)(m_4^2 + y_{L4}^2 f^2 - m_1^2)} {\rm Im}\, c_L\\
\rule{0pt}{1.75em}\displaystyle c_{\widetilde T,L} = -\sqrt{2} v \frac{y_{L1} y_{L4} m_4 f}{m_1(m_4^2 + y_{L4}^2 f^2 - m_1^2)} {\rm Im}\, c_L
\end{array}
\right.
\end{equation}
for the left-handed field interactions and
\begin{equation}
\left\{
\begin{array}{l}
\displaystyle c_{top,R} = - \sqrt{2} v \frac{y_{L4} y_{Lt} f}{m_4^2 + y_{L4}^2 f^2} {\rm Im}\, c_t\\
\rule{0pt}{1.75em}\displaystyle c_{T,R} = \sqrt{2} v \left[\frac{y_{L4} y_{Lt} f}{m_4^2 + y_{L4}^2 f^2} {\rm Im}\, c_t
- \frac{y_{L4} y_{L1} f}{m_4^2 + y_{L4}^2 f^2 - m_1^2} {\rm Im}\, c_R\right]\\
\rule{0pt}{1.75em}\displaystyle c_{\widetilde T,R} = \sqrt{2} v \frac{y_{L4} y_{L1} f}{m_4^2 + y_{L4}^2 f^2 - m_1^2} {\rm Im}\, c_R
\end{array}
\right.
\end{equation}
for the right-handed ones.

\begin{figure}
\centering
\includegraphics[width=.55\textwidth]{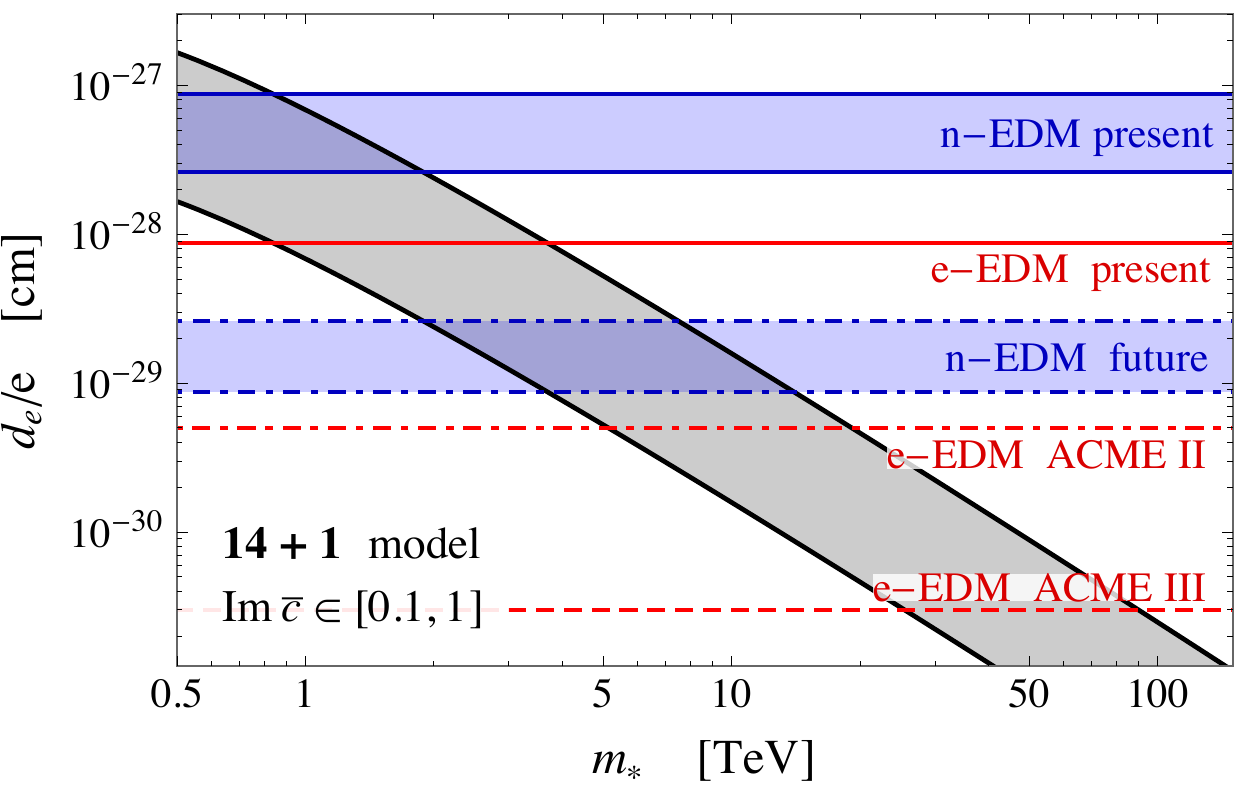}
\caption{Estimate of the bound on the lightest top partner mass in the $\bf 14 +1$ model with a fourplet and a singlet.
The gray band shows the estimate of the corrections to the electron EDM given in eq.~(\ref{eq:e-EDM_est_14+1})
for ${\rm Im}\,\bar c \in [0.1, 1]$. The solid red line shows the bound from the present electron EDM measurements, while
the dot-dashed and dotted ones show the expected future limits. The blue bands show the constraints from the
present and near-future neutron EDM measurements.
}\label{fig:bounds_14+1_full}
\end{figure}

Interestingly, all CP-violating couplings show a similar power-counting scaling, independently of the fact that they
originate from a $d$-symbol operator involving the $t_R$ or involving only top partners.
We generically expect $y_{L4} \sim y_{L1} \sim y_{Lt} \sim y_{top}$, $m_4 \sim m_1 \sim m_*$ and $c_L \sim c_R \sim c_t \sim 1$, so that all the couplings scale like $c \sim v f y_{top}^2/m_*^2$. As a consequence the contributions to the
Barr--Zee effects coming from the various $d$-symbol operators will be roughly of the same size.
Using these estimates we can easily derive the typical size of the contributions to the electron EDM as a function of
the top partners mass scale $m_*$,
\begin{equation}\label{eq:e-EDM_est_14+1}
\frac{d_e}{e} \sim \frac{e^2}{48 \pi^4} \frac{y_e}{\sqrt{2}} {\rm Im}\, \bar c\, \frac{y_{top}^2 v}{m_*^2} \log \frac{m_*^2}{m_{top}^2}\,.
\end{equation}
In the above formula we included a factor ${\rm Im}\,\bar c$, which encodes the typical size of the CP-violating
part of the $d$-symbol operator couplings. An analogous formula can be straightforwardly derived for the contributions
to the quark dipole moments.

In fig.~\ref{fig:bounds_14+1_full} we compare the estimate in eq.~(\ref{eq:e-EDM_est_14+1}) with the present
and projected future bounds from measurements of the electron and neutron EDM. To take into account possible accidental
suppressions we vary the factor ${\rm Im}\, \bar c$ in the range $[0.1, 1]$. One can see that the present bounds
can roughly test top partner masses of order $\textit{few}\;$TeV. The near-future improvements in the electron and
neutron EDM's can push the bounds in the range $5-10\;$TeV, while ACME III could test partners with masses
of order $40-100\;$TeV. We checked that the estimate in eq.~(\ref{eq:e-EDM_est_14+1}) is in good agreement with
the results obtained through a numerical scan on the parameter space of the model.

\subsection{The $\bf 5+5$ 2-site model}\label{sec:2-site}

As a second scenario we consider the $2$-site construction presented in refs.~\cite{Panico:2011pw,Matsedonskyi:2012ym} (see also ref.~\cite{DeCurtis:2011yx} for a similar set-up). This model is based on an extended set of global symmetries
that ensure the calculability of the Higgs potential. For definiteness we will focus on the scenario in which the  $q_L$ and $t_R$ fields are both elementary and are mixed with composite operators transforming in the fundamental representation of $\SO(5)$ (we thus dub this
set-up the `${\bf 5+5}$' model). This model can also be interpreted as a ``deconstructed'' version of the
MCHM$_5$ holographic scenario~\cite{Agashe:2004rs}.

The field content of the $\bf 5+5$ $2$-site model contains one set of composite top partners that transform as a fourplet and as a singlet under the unbroken $\SO(4)$ symmetry. The effective Lagrangian of the model can be written as
\begin{eqnarray}
{\cal L} & = & i \overline q_L \slashed D q_L + i \overline t_R \slashed D t_R + i \overline \psi_4 (\slashed D - i \slashed e) \psi_4 + i \overline \psi_1 \slashed D \psi_1 - \left(m_4 \overline \psi_{4L} \psi_{4R} + m_1 \overline \psi_{1L} \psi_{1R} + {\rm h.c.}\right)\nonumber\\
&& +\, \left(y_L f \overline q_L^{\bf 5} U \Psi + y_R f \overline t_R^{\bf 5} U \Psi - i c_L \overline \psi_{4L}^i \gamma^\mu d_\mu^i \psi_{1L} - i c_R \overline \psi_{4R}^i \gamma^\mu d_\mu^i \psi_{1R} + {\rm h.c.}\right)\,,\label{eq:5+5_Lagr}
\end{eqnarray}
where $\Psi = (\psi_4, \psi_1)$ denotes the $\SO(5)$ multiplet in the fundamental $\SO(5)$ representation built from the
$\psi_4$ and $\psi_1$ fields. Notice that the $\SO(4)$ symmetry would allow for four independent mixing terms
of the elementary $q_L$ and $t_R$ fields with the $\psi_4$ and $\psi_1$ multiplets. The structure in eq.~(\ref{eq:5+5_Lagr}) is dictated by the requirement of calculability of the Higgs potential.

All the parameters in the effective Lagrangian can in general be complex. By field redefinitions, three parameters can be made real, leaving $3$ physical complex phases. A convenient choice, which we will use in the following, is to remove the
phases from the elementary-composite mixings $y_L$ and $y_R$ and from one of the top partners mass parameters,
either $m_1$ or $m_4$. With this convention, the coefficients of the $d_\mu$-symbol operators remain in general complex.

Two free parameters can be chosen by fixing the top and Higgs masses. The top mass, at leading order in the $v/f$
expansion is given by
\begin{equation}\label{eq:top_mass}
m_{top}^2 \simeq \frac{1}{2} \frac{y_L^2 y_R^2 f^2 |m_4 - m_1|^2}{(|m_4|^2 + y_L^2 f^2)(|m_1|^2 + y_R^2 f^2)}v^2\,.
\end{equation}
The Higgs mass can be conveniently related to the masses of the top partners, namely~\cite{Matsedonskyi:2012ym} (see also ref.~\cite{Marzocca:2012zn})
\begin{equation}\label{eq:Higgs_mass}
m_h \simeq m_{top} \frac{\sqrt{2 N_c}}{\pi} \frac{m_T m_{\widetilde T}}{f} \sqrt{\frac{\log(m_T/m_{\widetilde T})}{m_T^2 - m_{\widetilde T}^2}}\,,
\end{equation}
where $N_c = 3$ is the number of QCD colors, while $m_T$ and $m_{\widetilde T}$ denote the masses of the
top partners with the quantum numbers of the top left and top right components respectively. The $T$ and $\widetilde T$
masses are approximately given by
\begin{equation}
m_T \simeq \sqrt{|m_4|^2 + y_{L}^2 f^2}\,,
\qquad
m_{\widetilde T} \simeq \sqrt{|m_1|^2 + y_{R}^2 f^2}\,.
\end{equation}
This relation (\ref{eq:Higgs_mass}) is valid with fair accuracy, $\sim 20\%$, and is only mildly
modified by the presence of additional heavier top partners.

Remarkably, eq.~(\ref{eq:Higgs_mass}) implies a tight relation between the mass of the lightest top partners and the Goldstone decay constant $f$, namely
\begin{equation}\label{eq:m_lightest_est}
m_{lightest} \lesssim \frac{\pi}{\sqrt{3}} \frac{m_h}{m_{top}} f \simeq 1.4\, f\,.
\end{equation}
Exclusion bounds on the top partner masses can thus be translated into lower bounds on the compositeness scale $f$.
The relation in eq.~(\ref{eq:m_lightest_est}) is saturated only if $m_T \simeq m_{\widetilde T} \simeq m_{lightest}$.
If the $T$ and $\widetilde T$ masses are significantly far apart, the lightest partner can be even a factor of $\sim 2$
lighter than the estimate in eq.~(\ref{eq:m_lightest_est}).

Let us now discuss the CP-violating effects.
We start by considering the properties of the Yukawa couplings. We saw that in the $\bf 14 + 1$ model,
all the mass parameters and elementary-composite
mixings can be made real by field redefinitions, therefore the Yukawa couplings alone can not generate CP-violating effects.
The situation is different in the $\bf 5+5$ set-up, in which one physical complex phase can not be removed from the
$y_{L,R}$ and $m_{4,1}$ parameters. In principle this could allow for CP-violating Yukawa couplings. Noticeably, in the
fermion mass eigenstate basis, only the off-diagonal Yukawa interactions can be complex, while the diagonal ones are
necessarily real. We will now present a general proof of this result that will allow us to identify the structural properties
from which it stems and the class of models for which it is valid.

The dynamics of the various resonances and their couplings with the Higgs can be encoded into a formal effective
Lagrangian obtained by integrating out all the top partner fields in the gauge interaction basis.
The only fields remaining
in this effective description are the elementary components $q_L$ and $t_R$.\footnote{This effective
description is analogous to the ``holographic'' effective Lagrangian in extra-dimensional models,
which is a function of the UV boundary values of the extra-dimensional fields~\cite{holography}.}
Notice that these fields have an overlap with
the whole set of mass eigenstates, thus they can describe any of them by just imposing the appropriate mass-shell condition. The effective Lagrangian contains operators with the generic form
\begin{equation}
i \overline q_L^{\bf 5} p^{2n} \slashed D q_L^{\bf 5}\,,
\qquad
i \overline t_R^{\bf 5} p^{2n} \slashed D t_R^{\bf 5}\,,
\end{equation}
which correct the kinetic terms of the $q_L$ and $t_R$ fields. These operators, however, are necessarily real, so they
do not give rise to CP-violating effects. The effective Lagrangian also contains a unique ``mass'' term, namely
\begin{equation}
\overline  m\, \overline q_L^{\bf 5} U t_R^{\bf 5} + {\rm h.c.}\,,
\end{equation}
which is the only invariant allowed by the symmetry structure of the model that does not contain derivatives.
This operator gives rise not only to the mass terms but also to the Yukawa couplings.

The $\overline m$ coefficient is in general complex. Nevertheless, when we redefine the fields to make the masses real,
we automatically remove all complex phases from $\overline m$. In such a way also the diagonal Yukawa couplings
are automatically made real. Notice that this result is true only in models in which a single ``mass'' invariant is present.
If multiple invariants are allowed, the Yukawa couplings are not ``aligned'' with the masses, thus making the masses real
in general does not remove the complex phases from the diagonal Yukawa couplings.
A scenario with multiple invariants can be obtained by embedding both the $q_L$ and the $t_R$ fields
in the $\bf 14$ representation of $\SO(5)$.

Since the diagonal Yukawa couplings are real, the only interactions that can generate CP-violating contributions
through Barr--Zee-type effects are the ones coming from the $d$-symbol operators.
Their explicit form at leading order in the $v/f$ expansion (using the convention in eq.~(\ref{eq:HGG_matrix_element})) reads
\begin{equation}
\left\{
\begin{array}{l}
\displaystyle c_{top,L} = -\sqrt{2} v f y_L^2 \frac{{\rm Im}[c_L (m_1 m_4^* + y_R^2 f^2)]}{(|m_4|^2 + y_L^2 f^2)(|m_1|^2 + y_R^2 f^2)}\\
\rule{0pt}{1.75em}\displaystyle c_{X_{2/3},L} = -\sqrt{2} v f y_R^2 \frac{{\rm Im}\,c_L}{|m_1|^2 + y_R^2 f^2 - |m_4|^2}\\
\rule{0pt}{1.75em}\displaystyle c_{T,L} = \frac{\sqrt{2} v f}{|m_4|^2 + y_L^2 f^2 - |m_1|^2 -y_R^2 f^2}\left[y_R^2 {\rm Im}\, c_L - y_L^2\frac{{\rm Im}[c_L (m_1 m_4^* + y_R^2 f^2)]}{|m_4|^2 + y_L^2 f^2}\right]\\
\rule{0pt}{1.5em}\displaystyle c_{\widetilde T, L} = - (c_{top,L} + c_{X_{2/3},L} + c_{T,L})
\end{array}
\right.
\end{equation}
for the left-handed field interactions and
\begin{equation}
\left\{
\begin{array}{l}
\displaystyle c_{top,R} = \frac{\sqrt{2} v f y_R^2}{|m_1|^2 + y_R^2 f^2} \left[\frac{{\rm Im}[c_R (m_1^* m_4 + y_L^2 f^2)]}{|m_4|^2 + y_L^2 f^2} + {\rm Im}[c_R m_1^*/m_4^*]\right]\\
\rule{0pt}{1.75em}\displaystyle c_{X_{2/3},R} = -\sqrt{2} v f y_R^2\frac{{\rm Im}[c_R m_1^*/m_4^*]}{|m_1|^2 + y_R^2 f^2 - |m_4|^2}\\
\rule{0pt}{1.75em}\displaystyle c_{T,R} = - \frac{\sqrt{2} v f}{|m_4|^2 + y_L^2 f^2 - |m_1|^2 -y_R^2 f^2}\left[y_L^2 {\rm Im}\, c_R - y_R^2\frac{{\rm Im}[c_R (m_1^* m_4 + y_L^2 f^2)]}{|m_4|^2 + y_L^2 f^2}\right]\\
\rule{0pt}{1.5em}\displaystyle c_{\widetilde T, R} = - (c_{top,R} + c_{X_{2/3},R} + c_{T,R})
\end{array}
\right.
\end{equation}
for the right-handed ones.

Interestingly, the dependence of the CP-violating coefficients on the elementary-composite mixings and on the masses
of the top partners is analogous to the one we found in the ${\bf 14 + 1}$ set-up. This result confirms that the
CP-violating effects in composite Higgs scenarios share some ``universal'' structure and are generically expected to
be sizable independently of the details of the model.

\begin{figure}
\centering
\includegraphics[width=.55\textwidth]{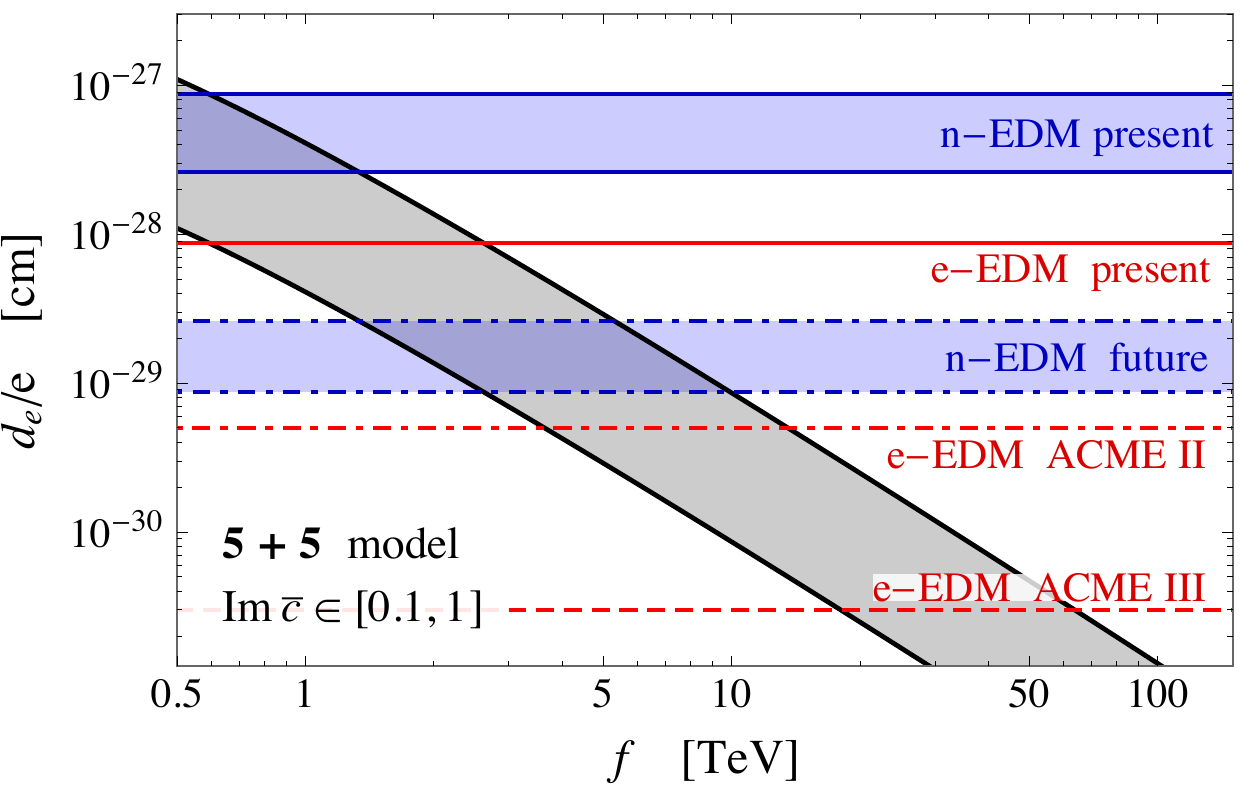}
\caption{Estimate of the bound on the compositeness scale $f$ in the $\bf 5 + 5$ model.
The gray band shows the estimate of the corrections to the electron EDM given in eq.~(\ref{eq:e-EDM_est_5+5})
for ${\rm Im}\,\bar c \in [0.1, 1]$. The solid red line shows the bound from the present electron EDM measurements, while
the dot-dashed and dotted ones show the expected future limits. The blue bands show the constraints from the
present and near-future neutron EDM measurements.
}\label{fig:bounds_5+5}
\end{figure}

Using the explicit expressions for the top mass in eq.~(\ref{eq:top_mass}), one finds that the elementary-composite
mixing parameters can be estimated as $y_L \sim y_R \sim y_{top} m_{lightest}/f$. Putting this result
together with the estimate in eq.~(\ref{eq:m_lightest_est}), we can express the corrections to the electron EDM
as a function of the compositeness scale $f$, namely
\begin{equation}\label{eq:e-EDM_est_5+5}
\frac{d_e}{e} \sim \frac{e^2}{48 \pi^4} \frac{y_e}{\sqrt{2}} {\rm Im}\, \bar c\, \frac{m_{top}}{1.4 f^2} \log \frac{(1.4 f)^2}{m_{top}^2}\,.
\end{equation}
This is a quite remarkable result, since it allows us to convert directly the bounds on dipole operators into constraints
on $f$. The numerical value of the estimate in eq.~(\ref{eq:e-EDM_est_5+5}) is shown in fig.~\ref{fig:bounds_5+5},
together with the experimental bounds. To allow for a certain amount of cancellation we varied the parameter
${\rm Im}\, \bar c$ in the range $[0.1, 1]$. The present data give bounds
$f \gtrsim 1\;$TeV. Near-future improvements in the electron and neutron EDM's will test $f \sim 5\;$TeV, while
the ACME III expected reach could probe $f \sim 50\;$TeV. Notice that these bounds are much stronger than the ones
coming from direct searches. As shown in ref.~\cite{Matsedonskyi:2015dns}, the LHC searches for top partners
can now exclude the $\bf 5 + 5$ model for $f \simeq 780\;$GeV, while the high-luminosity LHC program could
only slightly increase the bound up to $f \simeq 1.1\;$TeV.

It must be noticed that the estimate in eq.~(\ref{eq:e-EDM_est_5+5}) should be interpreted as a lower bound on the
corrections to the electron EDM. To derive it we assumed that the relation in eq.~(\ref{eq:m_lightest_est}) is saturated.
As we discussed before, this is true only if the $T$ and $\widetilde T$ masses are comparable. In generic
parameter space points the lightest partners can be even a factor $\sim 2$ lighter than the estimate, thus
leading to EDM contributions larger by a factor $\sim 4$. The presence of multiple CP-violating couplings can also
give rise to, small, additional enhancements. We verified by a numerical scan that the bounds in fig.~\ref{fig:bounds_5+5}
reproduce quite well the minimal constraints on $f$ as a function of the typical size of the complex phases.
They can thus be considered as robust constraints on the compositeness scale.

It is important to mention that the value of $\xi$ can be directly connected to the amount of fine-tuning~\cite{Panico:2012uw}.
In CH scenarios the $v/f$ ratio is not a free parameter, but rather a dynamical quantity fixed by the minimization of the radiatively-induced Higgs potential.
In generic parameter space points $\xi$ is expected to be of order one. Therefore, requiring a large separation between the Higgs vacuum expectation value and $f$ implies
a minimal amount of tuning of order $1/\xi$.\footnote{Note that additional sources of tuning can be present due to peculiarities
of the Higgs potential~\cite{Panico:2012uw}.} The constraints coming from the electron and neutron EDM's can thus be reinterpreted
as bounds on the minimal amount of fine-tuning in the $\bf 5+5$ $2$-site model. While $f \sim 1\;$TeV allows for a relatively low
tuning ($\xi \sim 0.1$), the future bounds are expected to test regions of the parameter space with a tuning significantly
below $1\%$.

To conclude the discussion about the $\bf 5+5$ model, we consider the contributions to the Weinberg operator.
Within the approximation in eq.~(\ref{eq:w_app}) we find
\begin{equation}\label{eq:W_5+5}
w \simeq \frac{g_s^3}{\sqrt{2} (4 \pi)^4} \frac{{\rm Im} (c_R - c_L) + \sqrt{2}\, {\rm Im} (c_R c_L^*)}{f^2}
\frac{|m_4|^2 - |m_1|^2}{|m_4 - m_1|^2}\,.
\end{equation}
Analogously to what we found for the $\bf 14 +1$ model (see section~\ref{sec:exp_bounds}), the top partners
contributions to the Weinberg operator do not decouple in the limit of heavy resonances.
The explicit result in eq.~(\ref{eq:W_5+5}) shows that, in addition to contributions linear in the $c_{L,R}$ parameters,
quadratic pieces are present. The latter come from diagrams involving two Higgs interactions coming from the
$d$-symbol operators. Notice that the above result is reliable only if $m_4 - m_1$ is not too small. In the limit
$m_4 = m_1$, the top mass vanishes (compare eq.~(\ref{eq:top_mass})) and the approximation in eq.~(\ref{eq:w_app})
is not valid.

To give an idea of the strength of the experimental bounds we fix the parameters by the relations $m_4 \sim m_1$
and $c_L \sim c_R$, moreover we set $\xi = 0.1$. The current bounds on the neutron EDM translate into
a bound $c_{L,R} \lesssim 1$, whereas the expected improved measurements will allow to probe $c_{L,R} \sim 0.1$.

\section{Conclusions}\label{sec:conclusions}

In this work we analyzed CP-violating effects induced by light top partners in composite Higgs scenarios.
We found that the main effects arise at two-loop level through Barr--Zee-type diagrams and generate sizable
contributions to the dipole moments of the electron and of the light SM quarks. Additional, although typically
subleading, contributions are induced for the purely-gluonic Weinberg operator.

Noticeably, in a large class of models, Barr--Zee effects arise exclusively from top partner interactions involving
the derivative of the Higgs field, namely $\partial_\mu h \overline \chi_i \gamma^\mu \chi_j$. The diagonal Yukawa
couplings, instead, are necessarily CP-conserving, thus not contributing to the light SM fermions dipole operators.
This result is valid in all models in which the effective Lagrangian contains only one invariant mass term for
the top quark (see section~\ref{sec:2-site}). Notice that this class of models is the most motivated one
from a flavor perspective, since a suppression of flavor-violating effects mediated by the Higgs~\cite{Agashe:2009di}
is also present. Without such feature very strong bounds from Higgs-mediated flavor-changing neutral currents would be present.

We found that the overall structure of the CP-violating effects, and in particular the dependence on the masses of the
top partners, is a rather universal feature and depends only mildly on the details of the model. The main contributions
to the electron and light quark dipole moments can be interpreted as a running effect. At the one-loop level the
top quark and its partners give rise to CP-odd contact interactions of the Higgs with the gauge fields (namely
$H^2 F_{\mu\nu} \widetilde F^{\mu\nu}$ with the photons and $H^2 G^a_{\mu\nu} \widetilde G^{a\,\mu\nu}$ with the gluons).
These operators, in turn, induce a running for the EDM's and CEDM's of the light SM fermions.
We explicitly computed how the contributions due to the top and its partners can be matched onto the CP-violating
Higgs contact interactions. In particular we found that running effects are always regulated at the top mass scale,
since the top contribution to the Higgs contact operators exactly balances the ones coming from the top partners.
Additional threshold contributions are found to be accidentally suppressed and numerically negligible.

In our analysis we focussed exclusively on the role of the top and its partners and we did not take into account
possible effects related to additional resonances. We also neglected the details of the flavor structure
both in the quark and in the lepton sectors. These aspects are expected not to spoil the overall
picture we described in this work. They could however have some impact on the bounds, which is worth exploring.
We leave this aspect for future investigation.

Although the CP-violating effects arise only at two-loop level, the present experimental bounds are tight enough
to give non-trivial constraints on the top partners masses. The strongest bounds come from the measurement of the
electron EDM, and can be used to probe top partners masses in the $\textit{few}\;$TeV range
(see figs.~\ref{fig:e-EDM_bounds} and \ref{fig:bounds_14+1_full}). Upgraded experiments
are expected to improve the bounds by one order of magnitude in the near future (ACME II) and by more than two
orders of magnitude at a later stage (ACME III), hence pushing the indirect exclusions for top partners well above
$10 - 20\;$TeV (fig.~\ref{fig:bounds_14+1_full}). Bounds from neutron EDM measurements are slightly
weaker than the ones from electron EDM, but could nevertheless test resonance masses in the $5-10\;$TeV range
in the near future.

In a large part of the parameter space of explicit models, the indirect bounds coming from the electron EDM
are competitive with the LHC direct searches for heavy vector-like quarks (see fig.~\ref{fig:comparison_with_direct}).
In particular CP-violating effects are induced by the same operators that control the
single-production vertices. In the absence of accidental cancellations or of accidentally small CP-violating phases,
the indirect bounds from CP violation tend to surpass the ones from single production searches.
The expected ACME II constraints will cover most of the LHC direct search reach even for complex phases as small as
$\textit{few}\,\%$.
ACME III could instead give constraints comparable with the direct ones achievable at future high-energy hadron colliders
such as FCC-hh with $100\;$TeV center of mass energy (see fig.~\ref{fig:comparison_with_direct_fcc}).

Interestingly, in specific scenarios such as the $\bf 5+5$ $2$-site model, the constraints fom CP-violating effects can be
translated into bounds on the Higgs compositeness scale $f$. While the present constraints are of order $f \gtrsim 1\;$TeV,
future improvements can push the bounds well above the $5 - 10\;$TeV range (see fig.~\ref{fig:bounds_5+5}). In these
scenarios the constraints on $f$ can also be translated into lower bounds on the amount of fine tuning.
For $f \sim 1\;$TeV the minimal fine-tuning is of order $5 - 10\%$, whereas it becomes $0.1\%$ for
$f \sim 10\;$TeV.

\section*{Acknowledgments}

We thank C.~Grojean, A.~Pomarol and A.~Wulzer for useful discussions. We also thank C.~Grojean and A.~Pomarol for
giving us comments on the manuscript.
M.R. is supported by la Caixa, Severo Ochoa grant program. G.P., M.R. and T.V. are supported by the Spanish Ministry MEC under grants FPA2015-64041-C2-1-P, FPA2014-55613-P and FPA2011-25948, by the Generalitat de Catalunya grant 2014-SGR-1450 and by the Severo Ochoa excellence program of MINECO (grant SO-2012-0234). G.P is also supported by the European Commission through the Marie Curie Career Integration Grant 631962 via the DESY-IFAE cooperation exchange.


\end{document}